\documentclass[longauth]{aa}
\usepackage{graphicx}
\usepackage{txfonts}
\usepackage{amssymb}
\usepackage{amsmath}
\usepackage{multirow}
\usepackage{rotating}
\usepackage{bm}
\usepackage{color}
\usepackage{hyperref}

\def\simlt{\lower.5ex\hbox{$\; \buildrel < \over \sim \;$}}
\def\simgt{\lower.5ex\hbox{$\; \buildrel > \over \sim \;$}}
\def\ltsim{\raise 2pt \hbox {$<$} \kern-1.1em \lower 4pt \hbox {$\sim$}}
\def\ltapprox{\raise 2pt \hbox {$<$} \kern-1.1em \lower 5pt \hbox {$\approx$}}
\def\gtsim{\raise 2pt \hbox {$>$} \kern-1.1em \lower 4pt \hbox {$\sim$}}
\def\gtapprox{\raise 2pt \hbox {$>$} \kern-1.1em \lower 5pt \hbox {$\approx$}}

\begin{document}

\title{Radio footprints of a minor merger in the Shapley Supercluster: From supercluster down to galactic scales}

\author{Venturi T.\inst{1}, Giacintucci S.\inst{2}, Merluzzi P.\inst{3}, Bardelli, S.\inst{4}, Busarello, G.\inst{3},
  Dallacasa, D.\inst{5,1}, Sikhosana, S.\,P.\inst{6,7}, Marvil, J.\inst{8}, Smirnov, O.\inst{9,10},
  Bourdin, H.\inst{11}, Mazzotta, P.\inst{11}, Rossetti, M.\inst{12}, Rudnick, L.\inst{13},
  Bernardi, G.\inst{1,9,10}, Br\"uggen, M.\inst{14}, Carretti, E.\inst{1},
  Cassano, R.\inst{1}, Di Gennaro, G.\inst{15,14}, Gastaldello, F.\inst{12}, Kale, R.\inst{16}, Knowles, K.\inst{9,10},
  Koribalski, B.\,S.\inst{17,18}, Heywood, I.\inst{19,20,10}, Hopkins, A.M.\inst{21}, Norris, R.\,P.\inst{18,17}, Reiprich, T.\,H.\inst{22},
  Tasse, C.\inst{23,24}, Vernstrom, T.\inst{25}, Zucca, E.\inst{4},
  Bester, L.\,H.\inst{10,9}, Diego, J.\,M.\inst{26}, Kanapathippillai, J.\inst{25}}

\institute
{
INAF -- Istituto di Radioastronomia, via Gobetti 101, I--40129, Bologna, Italy
\and
Naval Research Laboratory, 4555 Overlook Avenue SW, Code 7213, Washington, DC 20375, USA
\and
INAF -- Osservatorio Astronomico di Capodimonte, Via Moiariello 16, I--80131 Napoli, Italy
\and
INAF -- Osservatorio di Astrofisica e Scienza dello Spazio di Bologna - via Gobetti 93/3, 40129 Bologna, Italy
\and
Dipartimento di Fisica e Astronomia, Università di Bologna, Via Gobetti 93/2, 40129 Bologna, Italy,
\and
Astrophysics Research Centre, University of KwaZulu-Natal, Durban 4041, South Africa
\and
School of Mathematics, Statistics, and Computer Science, University of KwaZulu-Natal, Westville 3696, South Africa
\and
NRAO, PO Box 0, Soccoro, NM 87801, USA
\and
Department of Physics and Electronics, Rhodes University, PO Box 94, Makhanda 6140, South Africa
\and
South African Radio Astronomy Observatory, 2 Fir Street, Black River Park, Observatory, Cape Town 7925, South Africa
\and
Universit\'a di Roma Tor Vergata, Via della Ricerca Scientifica, I-00133 Roma, Italy
\and
IASF -Milano, INAF, Via A. Corti 12, I-20133 Milano, Italy
\and
Minnesota Institute for Astrophysics, University of Minnesota, 116 Church St. SE, Minneapolis, MN 55455, USA
\and
Hamburger Sternwarte, Universit\"at Hamburg, Gojenbergsweg 112, D-21029 Hamburg, Germany
\and
Leiden Observatory, Leiden University, PO Box 9513, 2300 RA Leiden, The Netherlands
\and
National Centre for Radio Astrophysics, Tata Institute of Fundamental Research, S. P. Pune University Campus, Ganeshkhind, Pune 411007, India
\and
Australia Telescope National Facility, CSIRO Astronomy and Space Science, P.O. Box 76, NSW 1710, Epping, Australia
\and
Western Sydney University, Locked Bag 1797, Penrith, NSW 2751, Australia
\and
Astrophysics, Department of Physics, University of Oxford, Keble Road, Oxford OX1 3RH, UK
\and
Department of Physics and Electronics, Rhodes University, PO Box 94, Makhanda, 6140, South Africa
\and
 Australian Astronomical Optics, Macquarie University 105 Delhi Rd, North Ryde, NSW 2113, Australia
\and
Argelander-Institut fu\"r Astronomie, Universita\"at Bonn, Auf dem H\"ugel 71, 53121 Bonn, Germany
\and
GEPI \& USN, Observatoire de Paris, CNRS, Universite Paris Diderot, 5 place Jules Janssen, 92190 Meudon, France
\and
Centre for Radio Astronomy Techniques and Technologies, Department of Physics and Electronics, Rhodes University, Grahamstown 6140, South Africa
\and
CSIRO Astronomy and Space Science, PO Box 1130, Bentley, WA, 6102, Australia
\and
Instituto de Fisica de Cantabria, CSIC-Universidad de Cantabria, E-39005 Santander, Spain
}

\date{Received ??-??-????; accepted ??-??-????}

\titlerunning{Radio footprints of a minor merger}
\authorrunning{T. Venturi}

\abstract
    {The Shapley Supercluster ($\langle z \rangle\approx0.048$) contains
      several tens of gravitationally bound clusters and groups, making it it is 
      an ideal subject for radio studies of cluster mergers.} 
    {We used new high sensitivity  radio observations to investigate the less energetic events of mass assembly
      in the Shapley Supercluster from supercluster down to galactic scales.} 
    {We created total intensity images of the full region between A\,3558 and A\,3562,
      from $\sim$\,230 to $\sim$\,1650~MHz, using ASKAP, MeerKAT and the GMRT, with sensitivities
      ranging from $\sim$6 to $\sim$\,100\,$\mu$Jy\,beam$^{-1}$.  We performed a detailed
      morphological and spectral study of the extended emission features, complemented
    with ESO-VST optical imaging and X--ray data from {\it XMM-Newton.}}
%
    {We report the first GHz frequency detection of extremely low brightness
      intercluster diffuse emission  on a $\sim$ 1 Mpc scale connecting a cluster and a group, namely: A\,3562 and the group SC\,1329--313. It is
      morphologically similar to the X-ray emission in the region.  
      We also found (1) a radio tail generated by ram pressure stripping
      in the  galaxy SOS\,61086 in SC\,1329--313; (2) a head-tail radio galaxy,
      whose tail is broken and culminates in a misaligned bar; (3) ultrasteep
      diffuse emission at the centre of A\,3558. Finally (4), we confirm the ultra-steep
      spectrum nature
      of the radio halo in A\,3562.}
%
    {Our study strongly supports
      the scenario of a flyby of SC\,1329--313 north of A\,3562 into the supercluster
      core.
      This event perturbed the centre of A\,3562, leaving traces of this interaction
      in the form of turbulence between A\,3562 and SC\,1329--313, at the
      origin of the radio bridge      and eventually affecting the evolution of individual supercluster galaxies by triggering
      ram pressure stripping. Our work shows that minor mergers can be spectacular and 
      that they have the potential to generate diffuse radio emission that carries important information
      on the formation of large-scale structures in the Universe.}
%


\maketitle

\section{Introduction}\label{sec:intro}

According to the hierarchical scenario for the formation of large-scale structures, galaxy clusters form as a consequence of mergers to reach and exceed masses on the order of 10$^{\rm 15}$ M$_{\odot}$. With a total energy output on the order of $10^{\rm 63}-10^{\rm 64}$\,erg, cluster mergers are the most energetic events in the Universe. The gravitational energy released into the cluster volume during such events deeply affects the dynamics of the galaxies, as well as the properties of the thermal and non-thermal (relativistic particles and magnetic field) components of the intracluster medium (ICM).

The close connection between the properties of the radio emission in galaxy clusters and their dynamical state is now an established result. In particular, giant radio halos and relics -- on a l Mpc-scale synchrotron steep-spectrum ($\alpha$ typically in the range [--1.4,--1.2] for S\,$\propto \nu^{\alpha}$) radio sources of $\mu$Jy~arcsec$^{-2}$ surface brightness and below \citep[see][for a recent observational overview]{vanWeeren19} -- are thought to be the result of particle re-acceleration due to turbulence and shocks induced in the
cluster volume during mergers and accretion processes \citep[see][for the most recent theoretical review]{Brunetti14}. 
While the details of the microphysics of these processes are still being investigated, the connection between radio halos, relics, and cluster mergers is supported by a number of observational results and statistical investigations as per \citep{Schuecker01,Cassano10,Kale15}, as well as studies on several individual clusters (e.g. A\,521, \cite{Bourdin13}; A\,1682, \cite{clarke19}; A\,2256, \cite{Ge20}).
The former results show that the number of radio halos is a steep function of the cluster mass and their detection in merging clusters increases considerably for masses M$\simgt 8\times10^{\rm 14}$ M$_{\odot}$ (see \cite{Liang00}, \cite{Buote01} and more recently \cite{Cuciti21}, \cite{Duchesne21a}, \cite{vanweeren21}). Moreover, it has been shown that less massive and/or less
energetic mergers may lead to the formation of radio halos with ultra-steep spectra, as
found in A\,521, whose spectrum with $\alpha\sim -1.9$ makes it a
prototype for ultra-steep spectrum radio halos \citep{Brunetti08,Dallacasa09}.

Most of our current knowledge on the connection between cluster mergers and diffuse cluster radio
sources is built upon observations of samples of intermediate to massive clusters, namely, M\,$\simgt 6-8\times10^{\rm 14}$ M$_{\odot}$, and characterised by mergers with a moderate mass ratio ($M_2/M_1\geq$\,1:4-1:5; i.e. \cite{Cassano16}).
However, the dominant (i.e. most frequent) process of mass assembly in the Universe is the accretion of systems in less extreme processes involving either smaller masses or mass ratios well below 1:4--1:5, namely, so-called 'minor mergers'.
The steep relation between the cluster mass (or X-ray luminosity) and the radio power for radio halos (i.e. \cite{Cuciti21}) and relics (i.e. \cite{deGasperin14}; \cite{Duchesne21b}), 
along with the limited sensitivity of radio interferometers has made the detection of such sources in less massive systems an extremely challenging task thus far and the observational signatures of minor mergers are still largely unexplored.

Against this backdrop, we can also see that a new era has just begun. The substantial leap forward in sensitivity and $u-v$ coverage offered by current interferometers -- such as
LOw Frequency ARray (LOFAR, \cite{vanHaarlem13},  Australian Square Kilometre Array Pathfinder (ASKAP, \cite{Hotan21}), MeerKAT (\citep{Camilo18}), and the upgraded Giant Metrewave Radio Telescope (uGMRT) -- is  broadening the observational parameter space considerably and opening a new window on phenomena that have been inaccessible thus far. Radio bridges of extremely low surface brightness are being discovered (i.e. A\,1758, \cite{Botteon20a}; A\,399--A\,401, \cite{Govoni19}) and, in addition, extremely long and twisted tailed radio galaxies as well as ultra-steep spectrum radio filaments are shown to be common inhabitants of galaxy clusters (i.e. A\,2034, \cite{Shimwell16}; Perseus, \cite{Gendron-Marsolais20}; A\,2255,
\cite{botteon20b}).

To take advantage of these new capabilities, we observed the central region of the Shapley Supercluster with ASKAP, MeerKAT, and GMRT in search of radio signatures of merger events in low- to intermediate-mass environments.
In this paper, we focus on the A\,3558 cluster complex in the central region of the Shapley Supercluster and we report on the findings in the whole region, which
includes the two clusters A\,3558 and A\,3562, and the two groups SC\,1327--312 and SC\,1329--313 between them. 
The layout of the paper is as follows. In Sect. \ref{sec:shapley}, we provide an overview of the region of the Shapley Supercluster  that is under investigation. In Sect. \ref{sec:obs}, we describe the radio observations and the data analysis. The radio images and spectral analysis are presented in Sect. \ref{sec:images}. Our findings are discussed in Sect. \ref{sec:disc}.  Conclusions are given in Sect. \ref{sec:conc}. 

Throughout the paper, we use the spectral power-law convention $S_\nu\propto\nu^{\alpha}$ and assume a cosmology based on H$_0=70$ km s$^{-1}$ Mpc$^{-1}$, $\Omega_{\rm m}=0.3,$ and $\Omega_\Lambda=0.7$.
At the average redshift of the Shapley Supercluster, that is, z=0.048, this gives a conversion factor of 0.928 kpc/arcsec and a luminosity distance of 210 Mpc.

\section{Shapley Supercluster in context}\label{sec:shapley}

The Shapley Supercluster \citep{shapley30} is one of the richest and most massive concentrations of gravitationally bound galaxy clusters in the local  Universe \citep[i.e.][]{Scaramella89,Raychaudhury89,Vettolani90,Zucca93}. It is located in the southern sky and lies behind the Hydra-Centaurus cluster.
Overall, the structure covers a redshift range $0.033 \simlt z \simlt 0.06$ 
\citep{Quintana95,Quintana97}, and has a mean redshift of $z\approx0.048$.

\begin{figure}[h!]
\includegraphics[scale=0.27,angle=-90]{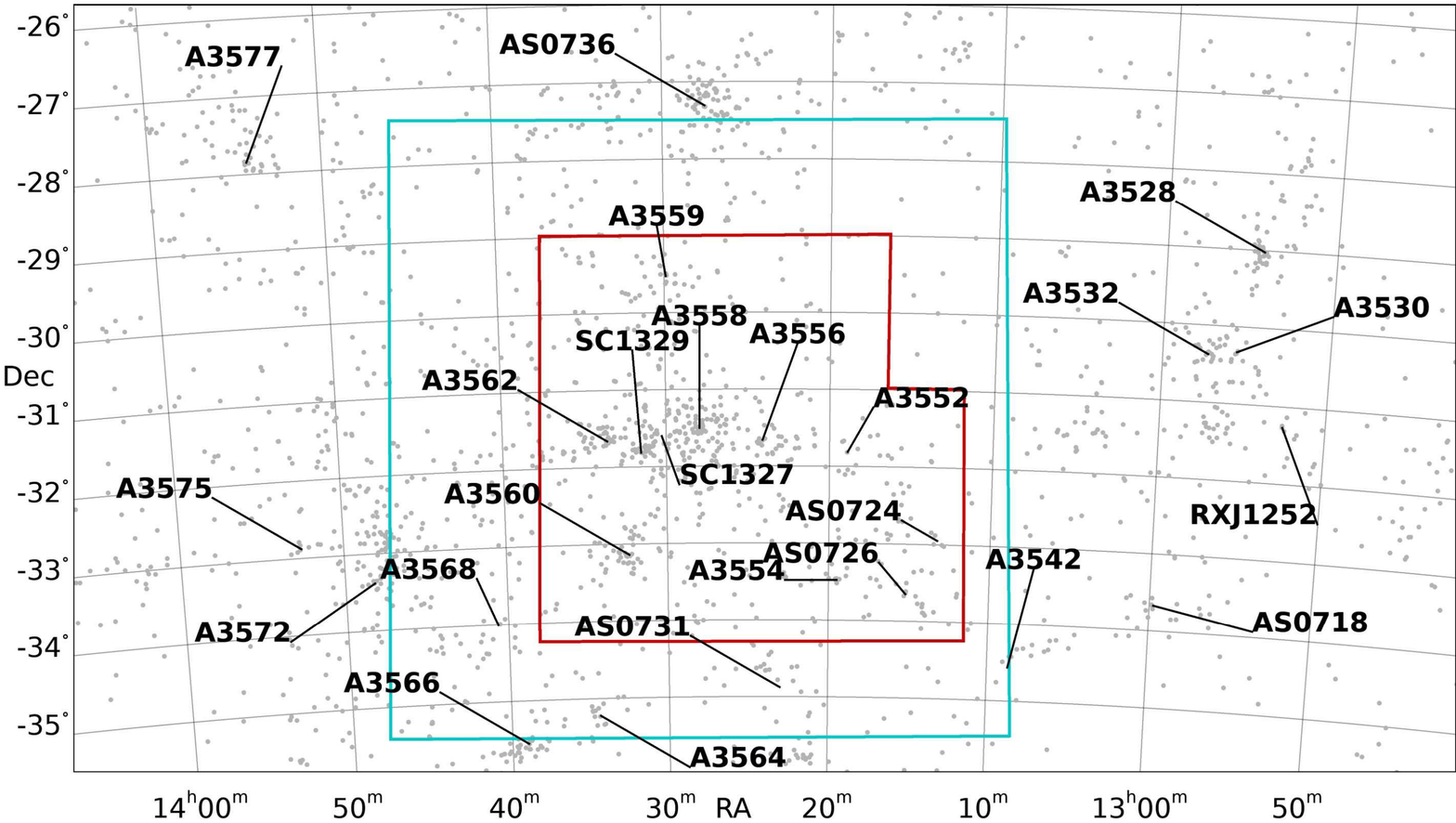}
\caption{Location of the clusters and groups in the Shapley Supercluster. The redshift information is from the 6dF Galaxy Survey  \citep[6dFGS,][]{Jones09}. The redshift range considered is 0.035--0.060. The cyan lines highlight the area covered by the ASKAP observations (Section 3.1), the red lines highlight the coverage of the ESO-VST optical imaging of ShaSS \citep{Merluzzi15}.}
\label{fig:fig1}
\end{figure}

Due to the very high overdensity and large number of galaxy clusters, and also thanks to its proximity, it is an ideal place to start investigating the effects of group accretion and cluster minor mergers, as is clear from the masses and bolometric X-ray luminosities of the individual clusters and groups, which range (estimated using 11 clusters, including A\,3552, see Fig. \ref{fig:fig1}) between M$_{500} \approx 0.4-9.8 \times 10^{14}$\,M$_{\odot}$ \citep{Higuchi20}\footnote{The dynamical masses presented in \cite{Higuchi20} have been transformed to M$_{500}$ by means of the software hydro\_mc(github.com/aragagnin/hydro\_mc) by \cite{Ragagnin20}.}, and L$_X \approx 0.4-6.7\times 10^{44}$ erg\,s$^{-1}$ \citep{deFilippis05} respectively, across  260\,Mpc$^2$ around the supercluster core.

The A\,3558 cluster complex is the centre of the Shapley Supercluster. It consists of a chain of three ACO \citep{Abell89} clusters (A\,3556, A\,3558 and A\,3562) and two smaller groups (SC\,1327--312 and SC\,1329--313).
It is extended for a projected length of about 7.5 h$^{-1}$ Mpc in the east-west direction
at an average redshift of $\langle z \rangle\approx0.048$. Figure \ref{fig:fig1} shows the clusters
and groups in the Shapley Supercluster. The cyan lines highlight the area covered by the
ASKAP observations (Section 3.1) and the red lines highlight the area of the Shapley Supercluster
Survey \citep[ShaSS][]{Merluzzi15,Mercurio15},
which includes optical imaging from ESO-VLT Survey Telescope (VST).
The main properties of the clusters studied in this paper are reported in Table~\ref{tab:info}.

Several studies in the infrared and visibile
\citep{Bardelli98a,Bardelli98b, Merluzzi15,Haines18,Higuchi20}, X--ray 
\citep{Markevitch97, Ettori00,Rossetti07,Finoguenov04,Ghizzardi10};
and radio (\cite{Venturi00}; \cite{Venturi03}, hereinafter V03; \cite{Giacintucci04} and \cite{Giacintucci05}, hereinafter G05; \cite{Venturi17a}, hereinafter V17) have provided observational evidence in
support of the idea that merging and accretion processes are taking place in the whole region between A\,3558 and A\,3562. This portion of the Shapley Supercluster has also been detected by Planck (see Fig. 1 in the Planck collaboration paper \cite{Planck14}).
On the western end of the chain, the cluster A\,3556 is rather faint in X--rays and shows a
dynamically relaxed state. The radio properties of the dominant galaxy in 
A\,3556 are consistent with its relaxed dynamical state as derived from the X--ray \citep{Venturi97,diGennaro18}.
%
%
%

%
%
%
\begin{table*}[h!]
  \caption[]{Properties of the clusters}
\begin{center}
\begin{tabular}{lcccccc}
\hline\noalign{\smallskip}  
Cluster & RA$_{\rm J2000}$ & DEC$_{\rm J2000}$ & v & $\sigma_{\rm v}$ & L$_{\rm X,bol}$ & M$_{\rm 500}$ \\
Name    & $^{\rm h,m,s}$ &  $^{\rm o,\prime,\prime\prime}$ & km s$^{-1}$ &  km s$^{-1}$ & 10$^{44}$ erg s$^{-1}$ & 10$^{13}$M$_{\odot}$ \\
\hline\noalign{\smallskip}
A\,3558       & 13 28 02 & --31 29 35 & 14500 $\pm$ 40 & 1007 $\pm$ 30 & 6.68  & 98$\pm$9 \\
SC\,1327--312 & 13 29 45 & --31 36 13 & 14800 $\pm$ 40 &  535 $\pm$ 20 & 1.27  & 20$\pm$2 \\
SC\,1329--313 & 13 31 36 & --31 48 45 & 13400 $\pm$ 50 &  373 $\pm$ 30 & 0.52  &  5$\pm$1 \\
A\,3562       & 13 33 47 & --31 40 37 & 14800 $\pm$ 55 &  769 $\pm$ 30 & 3.31  & 44$\pm$5 \\
\hline\noalign{\smallskip}
\end{tabular}
\end{center}
    {Notes: Cluster coordinates are taken from \cite{Merluzzi15}. Values for v
      and $\sigma_{\rm v}$ are taken from \cite{Haines18}, L$_{\rm X,bol}$ from \cite{deFilippis05}. M$_{\rm 500}$ is derived from M$_{\rm dyn}$ in \cite{Higuchi20} by means of the package hydro\_mc \citep{Ragagnin20}.}
\label{tab:info}
\end{table*}
%
%
%
%

\begin{table*}[h!]
\caption[]{Logs of the observations and image parameters}
\begin{center}
\begin{tabular}{lccccccccc}
\hline
\hline\noalign{\smallskip}
Cluster & Pointing centre & Array & Project & Obs. Date & $\nu^{\rm (a)}$  & $\Delta\nu^{\rm (b)}$ & t$_{\rm int}$
& FWHM & rms  \\
        & RA, DEC (J2000) &       &         &                &  MHz   &  MHz        &     h
& $^{\prime\prime},^{\circ}$     &  $\mu$Jy~beam$^{-1}$    \\
\hline\noalign{\smallskip}
A\,3558 & 13 25 51,  -31 03 05       & ASKAP & ESP\,20 & 19-Mar-19  & 887 & 288 & 11
&13.2$\times$10.4,  85.4 & $\sim$30--50 \\
A\,3558 & 13 27 54,  -31 29 32 & GMRT & 22\_039 & 30-Aug-12 & 306 & 32   & 7
& 14.0$\times$9.5, 18.4 & $\sim$ 60--100 \\
        &                              & GMRT & 22\_039 & 02-May-15 & 608 & 32   & 5
& 10.9$\times$5.6, 35.2 & $\sim$ 100     \\
        & 13 27 54,  -31 29 32       & MeerKAT & ${\rm (c)}$ & 04-Jul-18 & 1283 & 856 & 2
& 7.0$\times$5.9, 1.9 & $\sim$ 6 \\
SC\,1329--313 & 13 31 30,  -31 44 00   & GMRT & 30\_024 & 21-May-16 & 607 & 33 & 7
& 6.3$\times$3.2, 0     & $\sim$50 \\
              &                        & GMRT & 30\_024 & 22-May-16 & 233 & 33 & 7
& 24.4$\times$10.7, 26   & $\sim$500        \\
              & 13 31 08,  -31 40 23   & MeerKAT & ${\rm (d)}$ & 06-Jul-18 & 1283 & 856 & 8
& 7.3$\times$7.2, 84.1 & $\sim$ 6 \\
A\,3562 & 13 33 35,  -31 40 30 & MeerKAT & AO\,1 ${\rm (e)}$ & 07-Jul-19 & 1283 & 856 & 10
& 6.9$\times$6.5, 151.8 & $\sim$ 6 \\
\hline\noalign{\smallskip}
\end{tabular}
\end{center}
    {Notes: $^{\rm (a)}$ and $^{\rm (b)}$ refer to the central frequency and total bandwidth
      respectively.
      ${\rm (c)}$ The pointing is part of the MeerKAT scientific commissioning. ${\rm (d)}$
      The pointing is part of the MeerKAT Galaxy Clusters Legacy Survey. ${\rm (e)}$ 
      Observations obtained by the authors in response to the January 2019 MeerKAT Announcement of Opportunity.}
\label{tab:logs}
\end{table*}
%

\section{Observations and data analysis}\label{sec:obs}

The findings presented in this paper are based on five different datasets, collected with
GMRT, ASKAP, and MeerKAT, and cover a frequency range between 230 MHz to 1.65 GHz.
The details of our observations and final images are reported in Table~\ref{tab:logs}.
A description of the observational setup, calibration, imaging and mosaicing for each dataset
is provided in the following subsections, together with details on the data analysis
of the {\it XMM-Newton} observations used in the discussion.

\subsection{ASKAP}

The core of the Shapley Supercluster was observed with ASKAP as part of the EMU
Early Science programme \citep{Norris11,Johnston08} ESP\,20.
The observations (Table~\ref{tab:logs}) were carried out on 19 March 2019 (scheduling block 8140) with 35 out of the 36 antennas in the array. The antenna configuration provides a baseline coverage that ranges from $\sim$22~m to $\sim$6.4~km.
For this set of observations, the ASKAP correlator was used in pseudo-continuum mode to provide 288 channels, each with 1 MHz bandwidth, centred on 887 MHz.
The telescope was configured to produce 36 electronically formed beams arranged on the sky in a 6$\times$6 square grid; each individual beam covers an area of $\sim$1~deg$^2$, giving a total
instantaneous field of view of $\sim$ 31~deg$^2$ \citep{Hotan21}.
\\
At the beginning of each observation, the source B1934--638 was observed for 3 minutes at the centre of each beam to calibrate the delay, phase, bandpass, and flux-density, based on the model from Reynolds \citep{Reynolds94}.
The data reduction was performed using the ASKAPsoft software package and associated processing pipelines\footnote{https://www.atnf.csiro.au/computing/software/askapsoft/sdp/docs/current/}.
Images were produced using the w-projection algorithm to account for the w-term in the Fourier transform and with robust imaging weights to suppress the sidelobes of the point spread function.
The deconvolution process used the clean algorithm and included two Taylor terms to model the spectral variation of sources and several Gaussian scales to model the structure of extended sources.
Calibration and imaging were carried out separately for each beam, including one round of phase-only self-calibration, after which the images were combined with a linear mosaic to produce the final image.
A comparison between mosaics generated using simple 2d Gaussian beam models versus newer holography beam models using a large number of bright sources allows us to state that the residual amplitude calibration errors are on the order of 5\%.

\begin{figure*}[h!]
\centering
\includegraphics[scale=0.12]{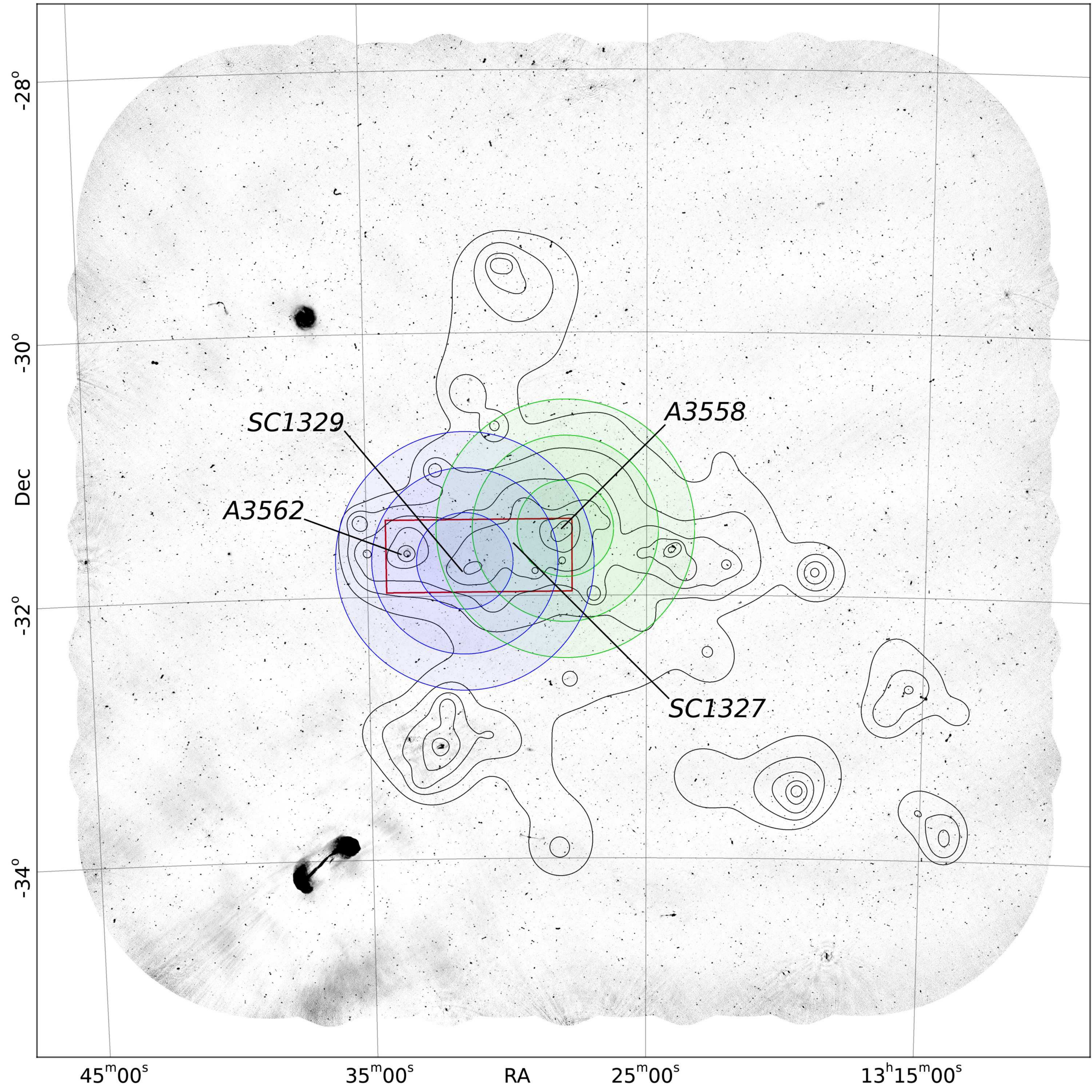}
\caption{Grey-scale view of the full field imaged by ASKAP at 887 MHz. The restoring beam is 
  $13.2^{\prime\prime}\times10.4^{\prime\prime}$, p.a. $85.4^\circ$.  The noise level ranges between 30
  and 50 $\mu$Jy~beam$^{-1}$ across the field. The radio galaxy PKS\,1333--33 is clearly visible
  in the bottom--left part of the field. The black contours show the number density of the
  galaxies
  at the redshift of the Shapley Supercluster. The contours correspond to 5, 10, 20, 40,
  80, 160 gal/Mpc$^2$ \citep{Haines18}.
  The region under study in this paper, going from A\,3562 to the central part of
  A\,3558 (east to west) is highlighted in the red rectangle.
  The green and purple circles show the GMRT pointings of the
    observations 22\_039 and 30\_024 respectively, and the full area covered by the primary beam.}
\label{fig:fig2}
\end{figure*}

The centre of our final mosaic, which is shown in Fig. \ref{fig:fig2}, is RA$_{\rm J2000}=13^h25^m50^s$, DEC$_{\rm J2000}=-31^{\circ}03^{\prime}05^{\prime\prime}$. The black contours in the image show the galaxy number density of the supercluster members \citep{Haines18}. The field of view of the ASKAP image covers most of the supercluster, as is clear from Fig. \ref{fig:fig1}.
The angular resolution is 13.2$^{\prime\prime}\times10.4^{\prime\prime}$ and the noise level is in the range 30--50\,$\mu$Jy~beam$^{-1}$.
As seen in Fig. \ref{fig:fig2}, the overall quality of the image is affected by the presence of the strong and extended radio galaxy  PKS\,1333--33 \citep{Killeen88,Condon21} located at the south--eastern edge of the field of view.
Moreover, artefacts in the shape of ripples in the north--south direction are also present.

\begin{figure*}[h!]
\vskip 0.8truecm
  \centering
{\includegraphics[scale=0.8]{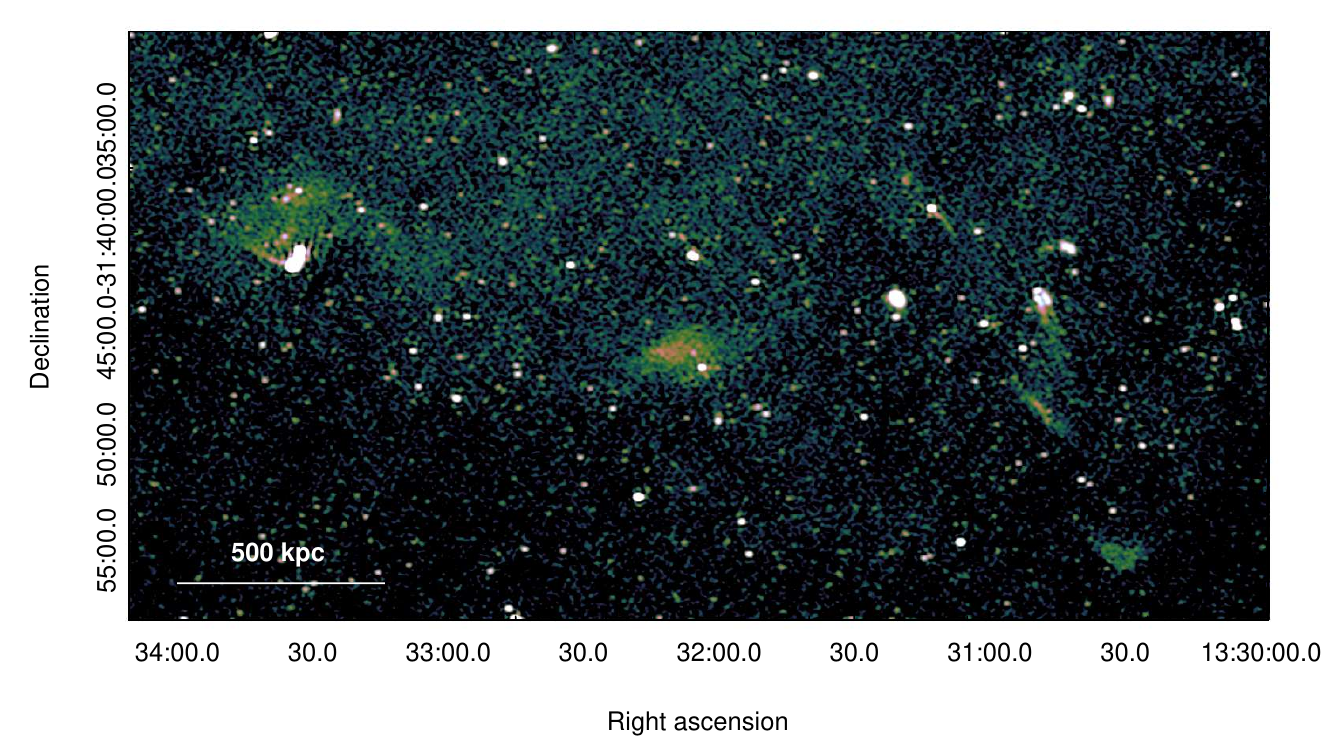}
\vskip 0.1cm
  \includegraphics[scale=0.78]{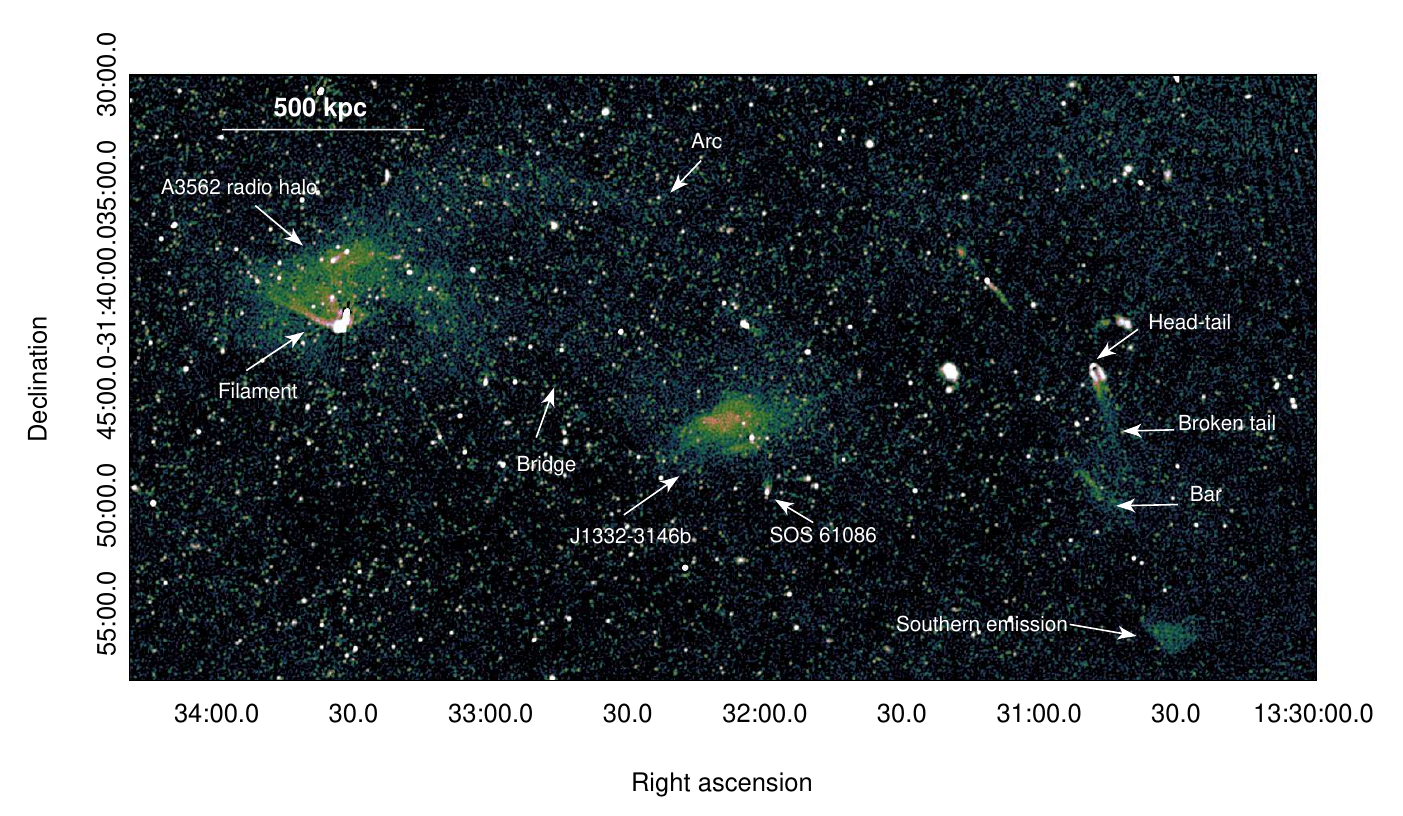}
\vskip 0.1cm
  \includegraphics[scale=0.70]{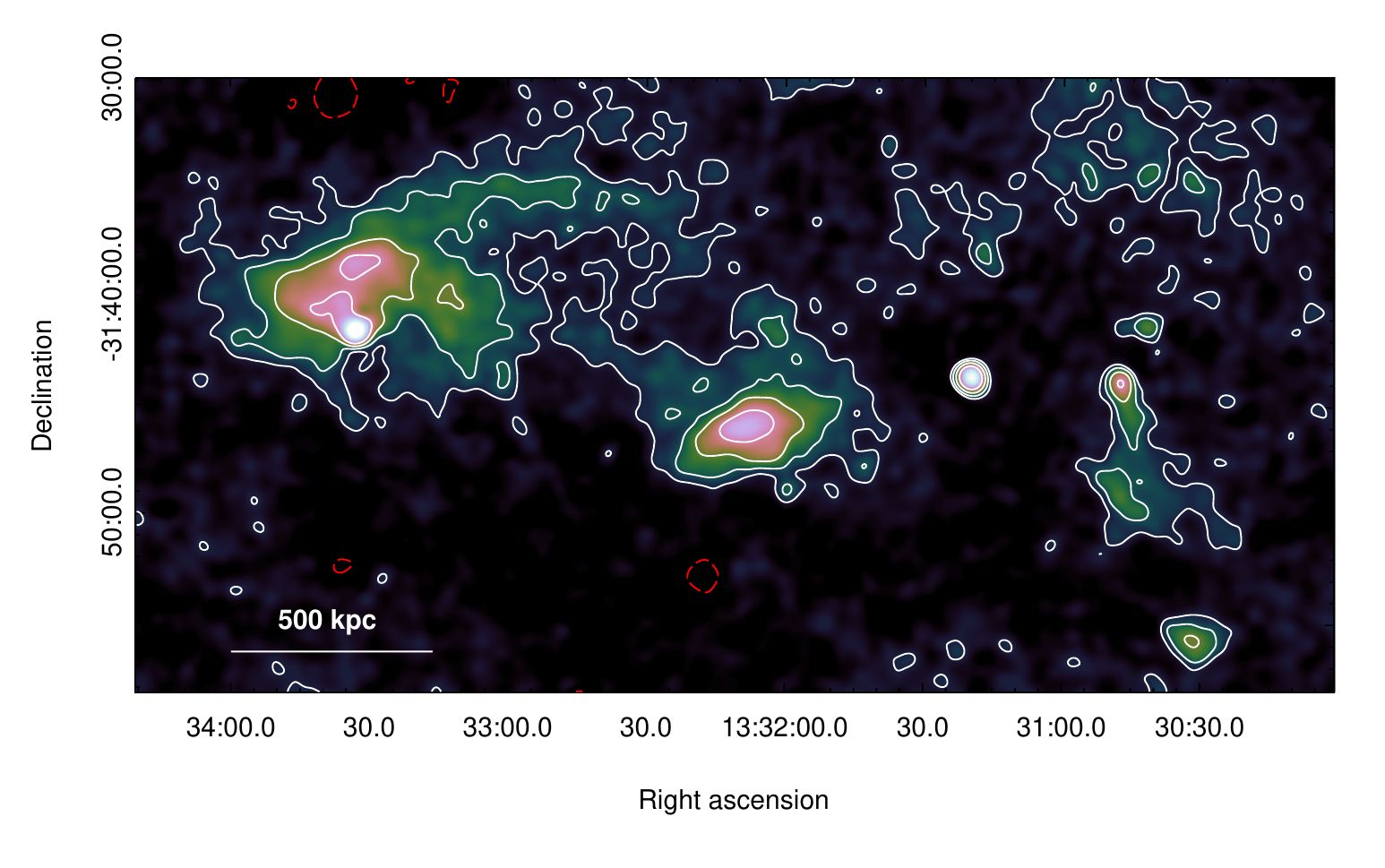}}
\caption{Colour scale of the 887 MHz ASKAP image zoomed into the area encompassing A\,3562 and
  SC\,1329--313 (upper panel). Restoring beam and noise level are as in Fig.~\ref{fig:fig2}.
  Central panel shows the 1.28 GHz MeerKAT image of the same region. The restoring beam is
  $6.9^{\prime\prime}\times6.5^{\prime\prime}$, p.a. $151.8^\circ$. 
  The average noise level is $\sim$ 6~$\mu$Jy\,beam$^{-1}$. The flux density colour scale is in the
  range -0.1--0.7 mJy. The labels highlight the features presented in this paper. Bottom panel shows the colour scale of the 1.28 GHz MeerKAT filtered image of the same region, convolved with
  a beam of $40^{\prime\prime}\times40^{\prime\prime}$ (see Sect. 3.2), with contours (dashed red
  for negative) of the same image overlaid. The contours
  are plotted at $\pm$0.1, 0.2, 0.4, 0.8 mJy\,beam$^{-1}$, the average noise level is $\sim$ 30~$\mu$Jy\,beam$^{-1}$.}
\label{fig:fig3}
\end{figure*}

The goals of our ASKAP Early Science Project are multi-fold. In this paper, we focus on the diffuse and extended emission associated with both the ICM and individual galaxies in the region between A\,3562 and SC1329--313 (zoomed in the upper panel of Fig. \ref{fig:fig3}) and in A\,3558 (upper panel of  Fig. \ref{fig:fig4}).
A detailed analysis of the radio galaxy population and the role of the environment by means of the
ShaSS dataset is in preparation.

%
%

%
%
%

\subsection{MeerKAT}
The MeerKAT observations come from various projects, as clarified in the notes to Table~\ref{tab:logs}.
Our observations on A\,3562, carried out in July 2019 (see Table~\ref{tab:logs}), were complemented with observations centered on SC1329--313 and A\,3558 as part of the MeerKAT Galaxy Cluster Legacy Survey \citep[MGCLS,][]{Knowles21} and of the MeerKAT scientific commissioning.

The July 2019 observatons were reduced as follows. 
We used the {\sc CARACal} pipeline\footnote{https://github.com/caracal-pipeline/caracal} \citep{Jozsa20a,Josza20b} for the initial data reduction. {\sc CARACal} orchestrates standard reduction packages into a single workflow. In this instance, it combined the {\sc Tricolour}\footnote{https://github.com/ska-sa/tricolour} flagger \citep{Hugo21} for RFI flagging, and standard {\sc CASA} tasks for reference calibration.
The Perley \& Butler \citep{Perley13} scale was used to set the flux density scale of B1934$-$638.
Whereas for the other standard MeerKAT flux density calibration source, PKS\,0408$-$65, we used a custom component-based field model provided in {\sc CARACal}, converted into model visibilities via the {\sc MeqTrees} package \citep{Noordam10}. After applying all the reference
calibrations, the data were averaged from the initial 4096 to 1024 channels, and imaged using the {\sc WSCLEAN} package \citep{Offringa14}.
We used Briggs weighting, robust = 0. We employed the joined-channel deconvolution and (4th order) polynomial fitting options of {\sc WSCLEAN} and made wideband images.  We imaged an area of $\approx 2\fdg2\times 2\fdg2$.
This was followed by a round of phase and delay self-calibration using the {\sc CubiCal}\footnote{https://github.com/ratt-ru/cubical} package \citep{Kenyon18}.

The data reduction of the pointings belonging to the MGCLS has been described in detail in \cite{Knowles21}.
The available products include in-band spectral index information $\alpha_{\rm 908~MHz}^{\rm 1656~MHz}$ within 36$^{\prime}$ from the pointing centres, and band-averaged Stokes I, Q, U, and V images.
\\
To double-check the consistency with the data analysis, the original $u-v$ data of the MGCLS pointings were passed through the pipeline developed by SARAO, with comparable results.
Both approaches produce images of extremely high quality, with rms values very close to the thermal noise,  rms $\sim$ 6~$\mu$Jy~beam$^{-1}$ at full resolution (see Table~\ref{tab:logs}).
To enhance the low-surface brightness emission in the field, we produced MGCLS images convolved to a resolution of 15$^{\prime\prime}$. The average noise at
this lower resolution is in the range $\sim$ 15-30~$\mu$Jy~beam$^{-1}$. The average residual amplitude calibration errors are of the order of 5\% for all MeerKAT datasets.

We finally produced images at the resolution of 40$^{\prime\prime}\times40^{\prime\prime}$ after removing the contribution of the individual sources. We followed two different approaches. We adopted both (1) the filtering method described in \cite{Rudnick02} and (2) the subtraction of individual galaxies from the $u-v$ plane  with further convolution of the residual emission \citep{Venturi07}.
The resulting images are consistent in the morphological details, but differ slightly in the flux densities integrated over large areas, the first method providing slightly higher values. Considering the very different approach of the two methods, we assume a conservative flux density uncertainty on the order of 20\%.
The images presented in this paper are those obtained with the filtering technique.

Our MeerKAT images of the region between A\,3562 and SC\,1329--313 are 
shown in the central and lower panel of Fig.~\ref{fig:fig3}. The emission 
at the centre of A\,3558 is shown in the lower panel of Fig.~\ref{fig:fig4}.

\subsection{GMRT}

The GMRT observations of project 22\_039 were pointed on A\,3558 (see Table~\ref{tab:logs}) and carried out at 306 MHz and 608 MHz, with a bandwidth $\Delta\nu$=32 MHz and 256 channels at both frequencies. The LL and RR polarisation were recorded in both bands.
The source 1311--222 was used as phase calibrator for both observations, while 3C\,286 was used as primary and bandpass calibrator.
A standard data reduction approach was carried out. In particular, a-priori calibration, initial flagging, and RFI excision were performed using {\it flagcal} \citep{Chengalur13}. Further editing, data averaging, and self-calibration were carried out using the NRAO Astronomical Image Processing System (AIPS). 
Direction-dependent calibration was successfully carried out at 306 MHz using the task PEELR in AIPS; however, this did not improve the quality of the 608 MHz image, since most of the strong sources in the field are resolved at the resolution of our observations. The final images, covering the full field of the primary beam, were produced over a range of angular resolutions, using different weighting schemes (Briggs parameters robust=0 and robust=+2) to highlight the details of the more compact features as well as the extended emission in the field. The final images have rms values in the range 60--100 $\mu$Jy~beam$^{-1}$ at 306 MHz and 608 MHz, respectively.

The region of the SC1329--313 group was observed with the GMRT on May 20 and 21, 2016 (project 30\_024) for 2.3 hours at 607 MHz and 5.6 hours at 233 MHz.
Both RR and LL polarisations were recorded.
The source 3C\,286 was observed at the beginning of each observation as a bandpass and absolute flux density calibrator.
The source 1311--222 was observed as a phase calibrator. All data were collected in spectral--line observing mode using the GMRT software correlator. A bandwidth of 33.3 MHz, 512 frequency channels, and a 4 second integration time were used at 610 MHz. At 235 MHz, the bandwidth was 16.7 MHz, divided in 256 channels, and the integration time was set to 8 seconds.
We used the Source Peeling and Atmospheric Modeling \citep[SPAM,][]{Intema09} pipeline to reduce the data using a standard calibration scheme consisting of bandpass and gain calibration and subsequent cycles of direction-independent self-calibration, followed by
direction-dependent self-calibration.
The flux density scale was set using 3C286 and the \citep{Scaife12}.
For the imaging, the final self-calibrated visibilities were first converted into a measurement set using the Common Astronomy Software Applications (CASA, version 5.6.0) and then imaged using WSClean 2.8.
We used the auto-masking and multi-scale algorithms \citep{Offringa17} and different values of the Briggs robustness parameter \citep[Briggs][]{Briggs95}, from $-0.5$ to $+0.5$, and $uv$ tapers.
The noise in the final images is of the order of 50~$\mu$Jy~beam$^{-1}$ at 607 MHz and 500~$\mu$Jy~beam$^{-1}$ at 233 MHz.
The residual amplitude calibration errors in the GMRT datasets are $\sim$5\% at 607 MHz, $\sim$ 8\% at 306 MHz, and $\sim$ 10\% at 233 MHz.
\\
Finally, primary beam correction was applied to all datasets with the task PBCOR in AIPS, following the GMRT guidelines\footnote{www.ncra.tifr.res.in:8081/$\sim$ngk/primarybeam/beam.html}. 
The full resolution of the final images are $14.0^{\prime\prime}\times9.5^{\prime\prime}$ at 306 MHz, $6.3^{\prime\prime}\times3.2^{\prime\prime}$ at 607 MHz and $24.4^{\prime\prime}\times10.7^{\prime\prime}$ at 233 MHz.
The primary beam corrected images are shown in the appendix.
Our datasets are sensitive to large angular scale emission, ranging from 17$^{\prime}$ to 44$^{\prime}$.

\subsection{{\it XMM-Newton} observations}\label{sec:xmm}

The sky area that covers the four constituents of the central region of the
Shapley Supercluster,  namely: A\,3558, A\,3562, SC\,1327--312, SC\,1329--313, and their connecting bridges, has been targeted by 11 observations performed with the three European Photon Imaging Cameras (EPIC). For more information, see Table \ref{tab:xmminfo}.

%
%
\begin{table}[h!]
  \caption[]{Logs of the EPIC {\it XMM-Newton} observations}
\begin{center}
\begin{tabular}{llll}
\hline\noalign{\smallskip}  
OBS-ID & RA   & DEC  & Field name \\
       & deg  & deg  &     \\
\hline\noalign{\smallskip}
0105261301  &  203.26196  &  --31.665306  &  A\,3562\_f1 \\
0105261401  &  202.96821  &  --31.827528  &  A\,3562\_f2 \\
0105261501  &  203.50542  &  --31.534639  &  A\,3562\_f3 \\
0105261601  &  203.13067  &  --31.652139  &  A\,3562\_f4 \\
0105261701  &  203.20183  &  --31.841722  &  A\,3562\_f5 \\
0105261801  &  203.57763  &  --31.711694  &  A\,3562\_f6 \\
0107260101  &  201.97975  &  --31.479778  &  A\,3558     \\
0601980101  &  202.31887  &  --31.711917  &  SC1327-312a  \\
0601980301  &  202.38137  &  --31.477667  &  SC1327-312b  \\
0651590101  &  202.65967  &  --31.749444  &  SC1329-313   \\
0651590201  &  202.38604  &  --31.612500  &  SC1329-312   \\
\hline\noalign{\smallskip}
\end{tabular}
\end{center}
\label{tab:xmminfo}
\end{table}
%
%

The surface brightness and temperature map and profiles presented in this paper (Sects. 4 and 5) have been obtained from the merging of these data sets into a composite event-list.

We sampled this event-list at an angular resolution of 6.5$^{\prime\prime}$ and an energy dependent spectral resolution in the range of 15-190 eV.
We associated a background noise model to the resulting event cube  and an effective area that follow the same angular and spectral sampling. The background noise model includes sky (Cosmic X-ray Background, two component Galactic Transabsorption Emission, see \cite{Kuntz00} and instrumental (soft proton and particle induced) components whose spatial and spectral templates have been jointly normalised with the hot gas emission outside the brightest regions of the Shapley Supercluster core (SSC).
Spectral analyses assume a hot gas emission that follows the Astrophysical Plasma Emission Code \citep[APEC,][]{Smith01} and an X-ray absorption calibration using an average Galactic neutral hydrogen density column of N$_{HI}=3.77 10^{20}$\,cm$^{-2}$, which we extracted in the SSC area from a map of the 21 cm emission released by the
Leiden/Argentine/Bonn Galactic HI survey 
\citep{Kalberla05}.

The surface brightness was obtained with a  wavelet analysis of photon images that
we corrected for spatial variations of the effective area and background model. Photon images have been denoised via the 4-$\sigma$ soft-thresholding of variance-stabilised wavelet transforms \citep{Zhang08,Starck09}, which are especially suited to processing low photon counts.
The image analyses include the inpainting of detected point-sources and the spatial adaptation of wavelet coefficient thresholds to the spatial variations of the effective area.

We computed the temperature map of the whole field using a spectral-imaging algorithm that combines likelihood estimates of the projected hot gas temperature with a B2-spline wavelet analysis. As detailed in \cite{Bourdin08} temperature log-likelihoods are first computed from spectral analyses performed within square bins of various angular resolutions, then convolved with analysis kernels that allows us to derive B2-spline wavelet coefficients and their expected fluctuation. We used such coefficients to derive a wavelet transform that typically analyse projected
temperature features of apparent size in the range of [0.2,2] arcmin, and we reconstructed a de-noised temperature map from a 3-$\sigma$ thresholding of the
wavelet coefficients.

%
\begin{figure}[h!]
\centering
{\includegraphics[scale=0.35]{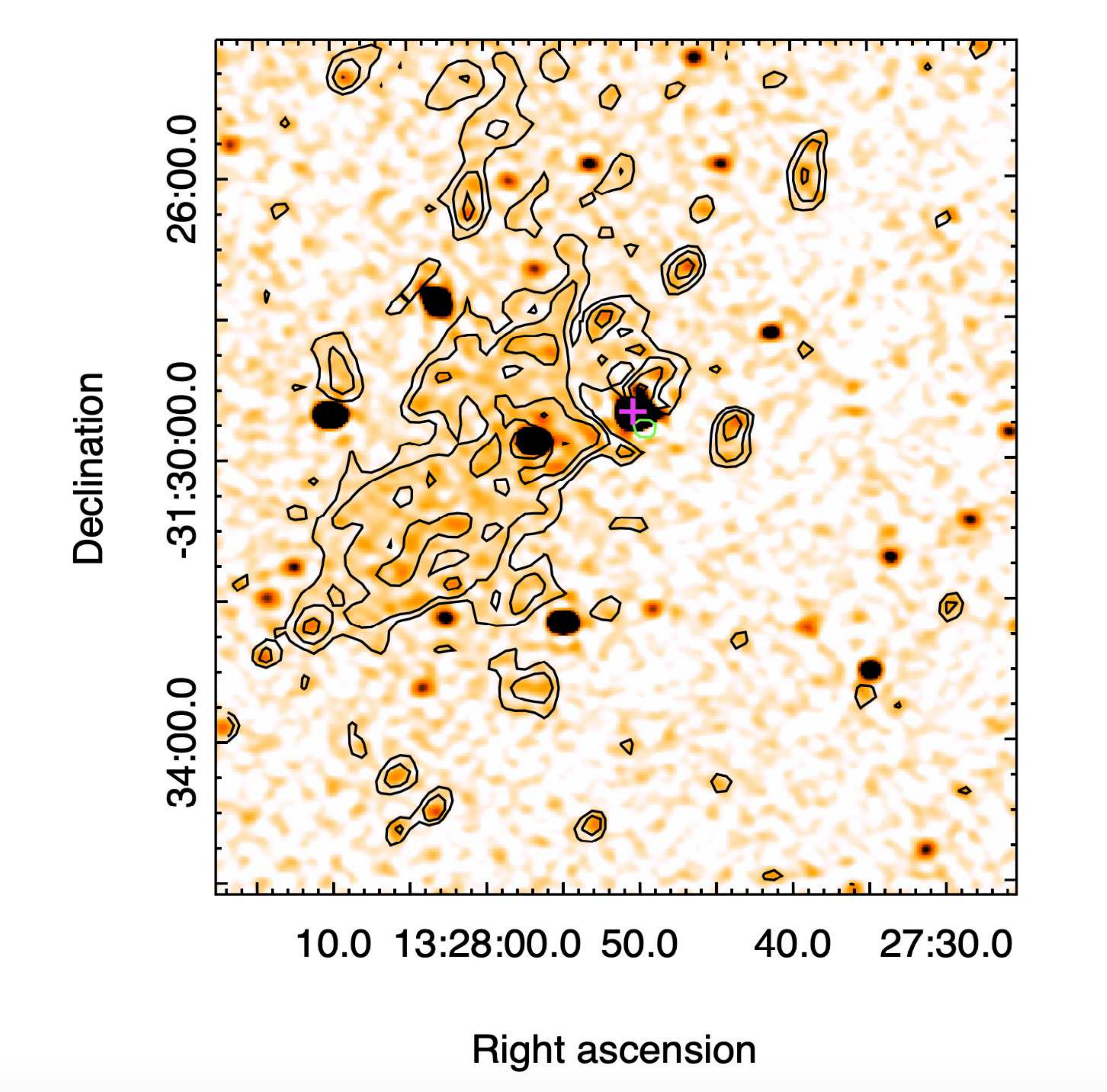}
\vskip 0.2truecm
\includegraphics[scale=0.35]{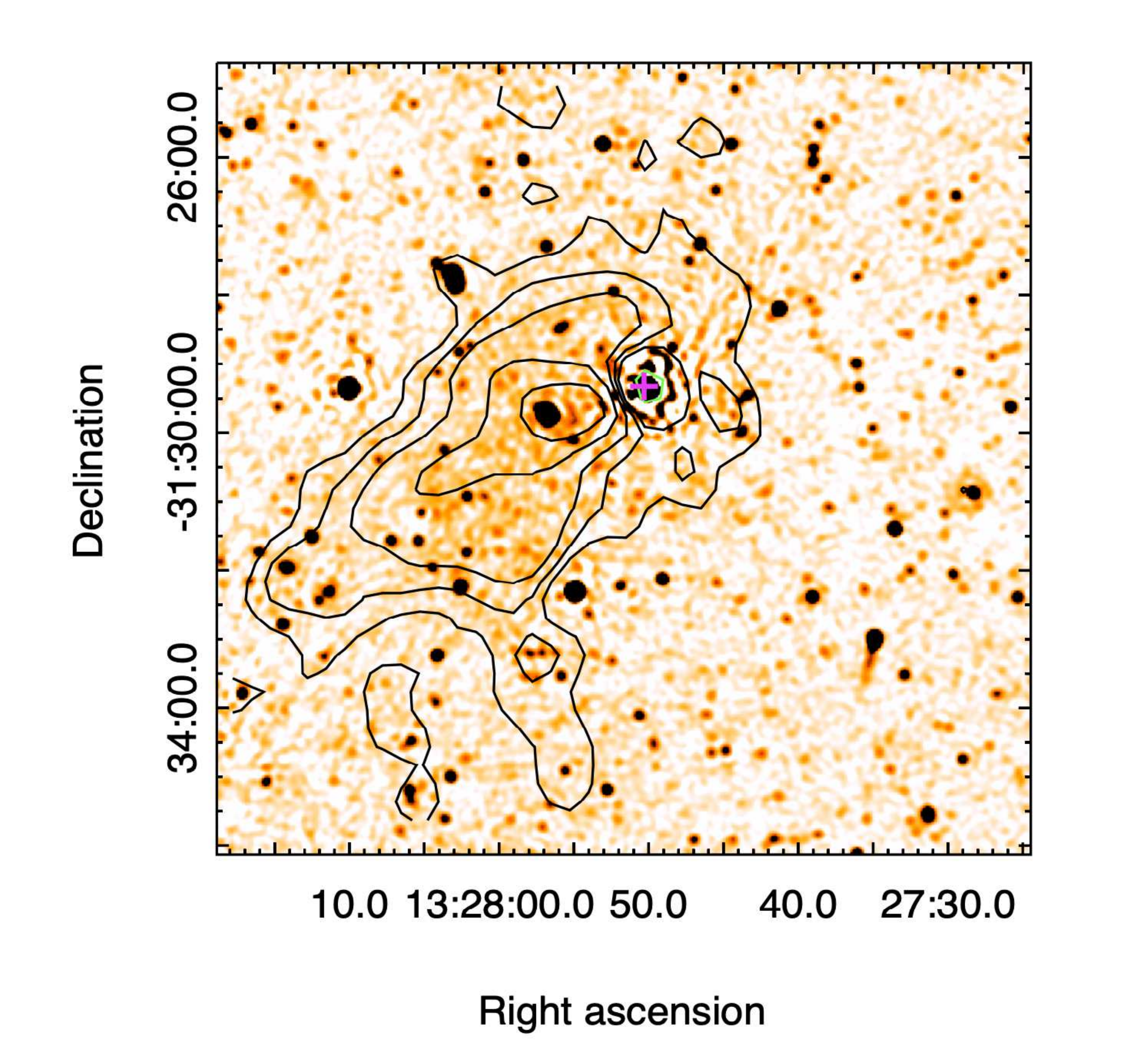}}
\caption{Radio halo in A\,3558. In both panels the magenta cross shows the
  position of the brightest cluster galaxy (BCG). Upper panel: ASKAP image at 887 MHz in colour
  (same resolution and rms as in the upper panel of Fig. \ref{fig:fig3}). The black contours
  show the emission of the halo after subtraction of the embedded point sources
  (see Sect. 4.7). The resolutions is $25.1^{\prime\prime}\times20.9^{\prime\prime}$,
  in p.a. $131.5^{\circ}$ and the rms is $\sim$35\,$\mu$Jy~beam$^{-1}$. Contour levels start at
  $\pm$0.15 mJy~beam$^{-1}$ and are spaced by $\sqrt2$ (negative contours in green). 
  Lower panel: MeerKAT image at 1.283 GHz in colour (same resolution and rms as in the
  central panel of Fig. \ref{fig:fig3}). The black contours show the emission after
  subtraction of the point sources (see Sect. 4.7)
  convolved with a restoring beam of $40.9^{\prime\prime}\times40.4^{\prime\prime}$,
  in p.a. $144.6^{\circ}$. The rms is $\sim$35\,$\mu$Jy~beam$^{-1}$. Contour levels start at
  $\pm$0.125\,mJy~beam$^{-1}$ and are spaced by $\sqrt2$ (negative contours in green).}
  \label{fig:fig4}
\end{figure}
%

\section{Radio images and spectral analysis}\label{sec:images}

\begin{figure*}[h!]
\centering
\includegraphics[scale=0.5]{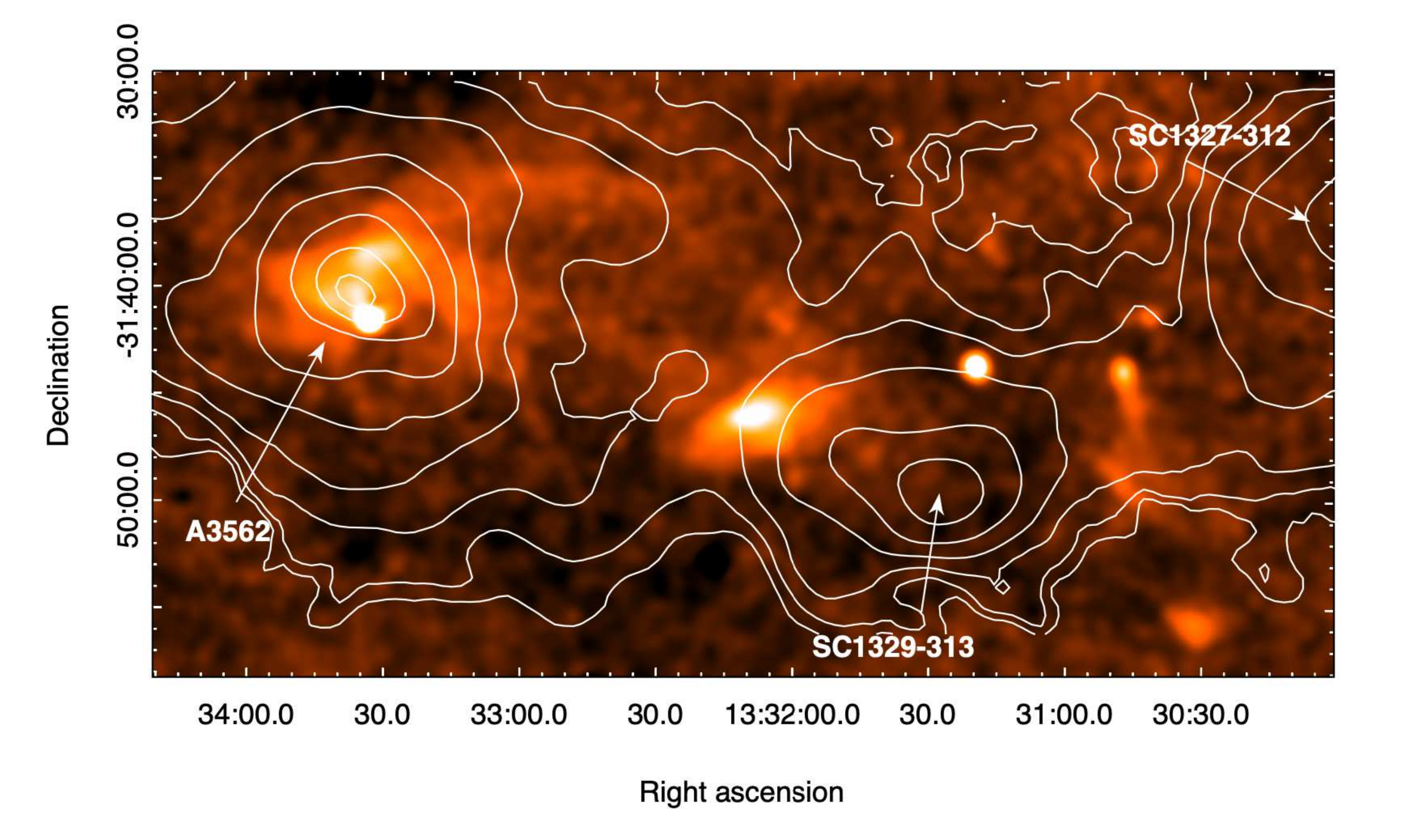}
\caption{MeerKAT 1.283 GHz colour scale of the diffuse emission
  (same field as Fig. \ref{fig:fig3}). The restoring beam is 
  $40^{\prime\prime}\times40^{\prime\prime}$. The contours of the
  {\it XMM-Newton} surface brightness (see Sect. 3.4)
  are overlaid in white and are spaced by a factor of 2.}
\label{fig:fig5}
\end{figure*}

Figures \ref{fig:fig3} and \ref{fig:fig4} clearly show that the
central region of the Shapley Supercluster is characterised by several
features of diffuse radio emission. Moreover, Fig. \ref{fig:fig5} shows the diffuse
emission in the region between A\,3562 and SC1329--313 with the thermal
emission detected by {\it XMM-Newton} overlaid in contours.

Going from east to west we identify (see the labels in the central panel of
Fig. \ref{fig:fig3}):
\begin{itemize}

\item[(a)] the well-known radio halo in A\,3562 (V03 and G05);

\item[(b)] diffuse emission of very low surface brightness, labelled as 'arc' and
  `bridge', detected here for the
  first time, which connects the radio halo in A\,3562 and  the radio source
  J\,1332--3146a in SC\,1329--313;
 
\item[(c)] the diffuse radio emission J\,1332--3146a, first imaged in G05;

\item[(d)] a resolved tailed emission just south of J\,1332--3146a (galaxy SOS\,61086)
  first imaged here;
  
\item[(e)] the head-tail first noticed in V17; 

\item[(f)] a faint extended emission just south of the head-tail, labelled as 'bar';

\item[(g)] a diffuse patch of emission $175^{\prime\prime}$ across, of unknown origin,
  labelled as `Southern emisson';
  
\item[(h)] faint diffuse emission at the centre of A\,3558, detected for the
  first time (Fig. \ref{fig:fig4}).

\end{itemize}

We will describe all these features and provide their observational parameters
in the next subsections. Their origin will be discussed in Sect. 5.

%
%

\subsection{Radio halo in A\,3562}\label{sec:halo}

The radio halo in A\,3562 was first imaged with the VLA at 1.4 GHz in V03 and further studied at several radio frequencies and in the X--ray band in G05. With a 1.4 GHz power of 1.14$\times10^{23}$ W~Hz$^{-1}$, it is a relatively low-power radio halo, which well fits the radio power--cluster mass correlation for galaxy clusters \citep{Cassano13}. V03 and G05 reported a spectral index of the order of $\alpha_{\rm 843\,MHz}^{\rm 1400\,MHz}\sim$\,--2.
This value of $\alpha$ is considerably steeper than usually found, that is, [--1.4,--1.2], placing this among the ultra-steep spectrum radio halos, first identified as a particular class of radio halos a few years later in \citep{Brunetti08}.

\begin{figure}[h!]
\centering
\includegraphics[scale=0.5]{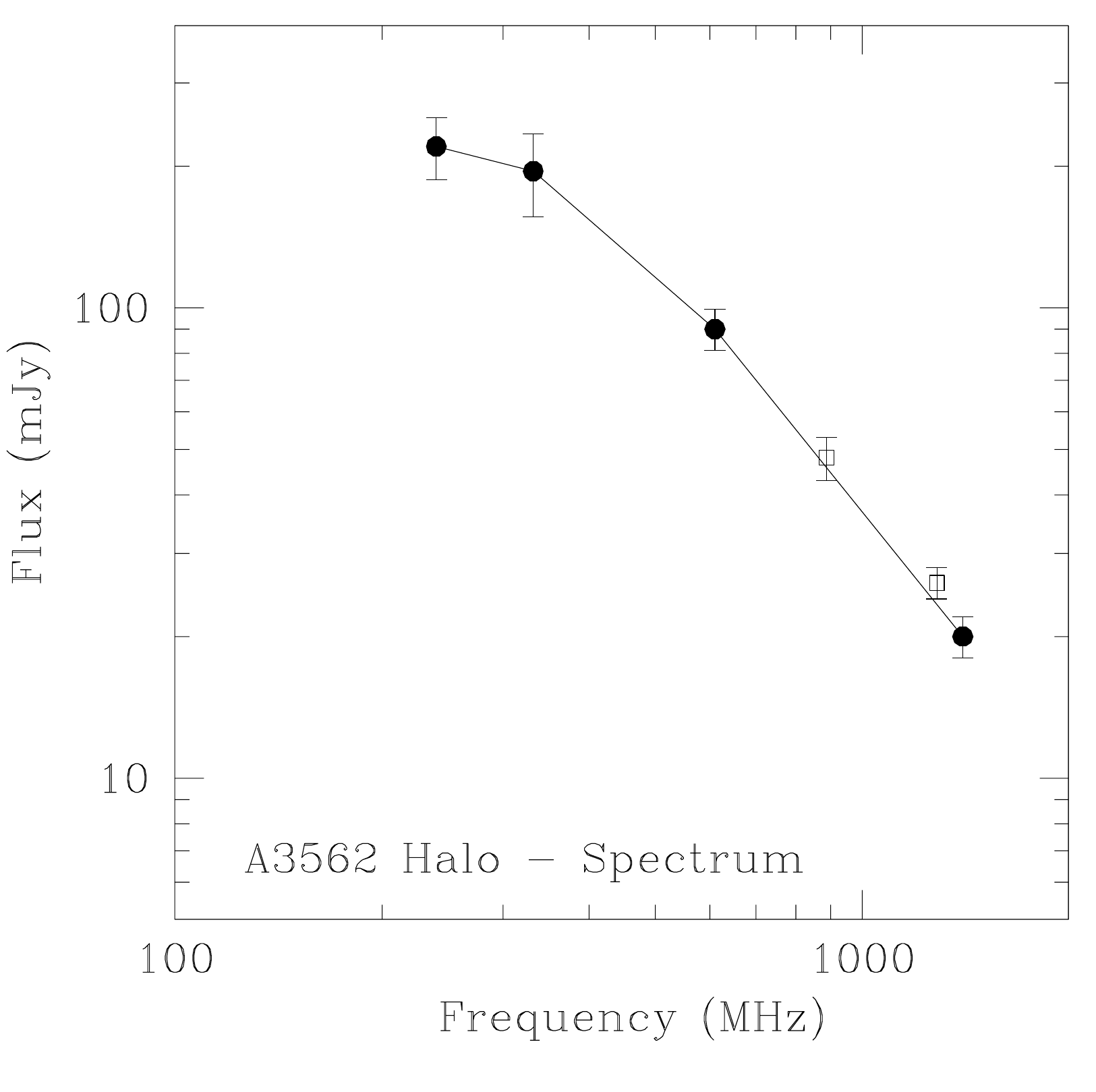}
\caption{Radio spectrum of the A\,3562 radio halo, obtained using the data in
  Table \ref{tab:fluxhalo}. Filled circles show the data published in V03 and G05,
  open squares are the new ASKAP and MeerKAT data presented here.}
\label{fig:fig6}
\end{figure}

Our MeerKAT and ASKAP images (Fig. \ref{fig:fig3}) show that this radio halo  extends further westwards towards the radio source J\,1332--3146a in SC\,1329--313 (see Sect. 4.3).
Its most noticeable new feature is the bright ridge of emission, labelled `filament' (see Fig. \ref{fig:fig3}, central panel), delimiting the south--eastern edge. Moreover, radio emission is clearly detected for the first time with MeerKAT south of this filament.
The surface brightness of the radio halo is of the order of 0.5\,$\mu$Jy/arcsec$^2$ and drops to  0.23\,$\mu$Jy/arcsec$^2$ in the newly detected region south of the filament.
A more detailed analysis of this feature and a revised discussion of the origin
of this radio halo will be presented in a separate paper (Giacintucci et al. in prep.).
Here, in Table \ref{tab:fluxhalo}, we report the flux density measurements of
the radio halo, including the new ones, after removal of the embedded point sources.
The updated radio spectrum is shown in Fig. \ref{fig:fig6}.
The MeerKAT and ASKAP flux density values were obtained
integrating over the same area imaged in V03 and G05 for a consistent
comparison. The new values perfectly align with the previous measurements
and suggest that the spectral index has a value
$\alpha^{\rm 1283~MHz}_{\rm 332~MHz}= -1.5\pm0.2$.  
The last two datapoints suggest a consistent steepening above 1283 MHz
(Giacintucci et al. in prep.), in agreement with our earlier work.

The discovery of the 
ultra--steep spectrum radio halo in the massive cluster A\,521 \citep{Brunetti08} led to the detection of a number of radio halos with ultra--steep spectrum \citep[see][and references therein]{vanWeeren19}. It is nowadays thought that such sources are related either to merger events involving clusters of small mass (M $\simlt$~6~$\times$10$^{14}$M$_{\odot}$) or to off--axis mergers \citep{Cassano12}.
Both the masses of A\,3562 and SC\,1319--313 (see Table \ref{tab:info}) and the X--ray properties of this region are consistent with the
presence of an ultra--steep spectrum of the radio halo in A\,3562.

%
%

%
%
\begin{table}[h!]
  \caption[]{Flux density values of the radio halo in A\,3562 }
\begin{center}
\begin{tabular}{cccc}
\hline\noalign{\smallskip}  
$\nu$ & Array & Flux & Ref \\
MHz   &       & mJy  &     \\
\hline\noalign{\smallskip}
1400 & VLA     &  20$\pm$ 2 & G05 \\
1283 & MeerKAT &  26$\pm$ 2 & This work \\
887  & ASKAP   &  48$\pm$ 5 & This work \\
610  & GMRT    &  90$\pm$ 9 & G05 \\
332  & GMRT    & 195$\pm$39 & G05 \\
240  & GMRT    & 220$\pm$33 & G05 \\
\hline\noalign{\smallskip}
\end{tabular}
\end{center}
\label{tab:fluxhalo}
\end{table}
%
%

\subsection{Diffuse radio emission connecting A\,3562 and the group SC\,1329--313}

The most striking result of our observations is the discovery of extended, very low surface-brightness emission in the region connecting A\,3562 and SC\,1329--313, as highlighted in the bottom  panels of  Fig. \ref{fig:fig3} and in Fig. \ref{fig:fig5}.
It is the first time that diffuse emission between a cluster and a group is detected at GHz frequencies.
Figure \ref{fig:fig5} shows a remarkable correlation between the morphological details of this feature and the X-ray emission as imaged by {\it XMM-Newton}.

The possible existence of a radio bridge connecting A\,3562 and the source J\,1332--3146a in SC\,1329--313 at 1.4 GHz was reported in G05, however, the sensitivity of those observations was not good enough for a robust claim. Similarly, the GMRT observations presented here are not adequate to reveal this feature, due to the pointing centres and limited sensitivity.
Our MeerKAT images now clearly detect the radio bridge, which extends from the western end of the radio halo in A\,3562 (labelled 'filament' in Fig. 9 in G05) to the eastern extension of J\,1332--3146a, covering a distance of $\sim$1 Mpc in projection.
Moreover, we detect an arc-like feature which connects A\,3562 and SC\,1329--313 from the north. The arc is very well imaged with MeerKAT, while it is less prominent in the ASKAP image, due to the presence of residual
ripples (see Sect. 3.1 and Figs. \ref{fig:fig2} and \ref{fig:fig3}).

We integrated the flux density of this emission over the image presented in the bottom panel of Fig. \ref{fig:fig3} and in Fig. \ref{fig:fig5} after primary beam correction (and removing the contribution of the radio halo in A\,3562), and obtained S=61$\pm$6mJy for the total of the diffuse emission (including both the bridge and arc) filling the region between A\,3562 and J\,1332--3146a, which corresponds to a radio power P$_{\rm (1.283~GHz)}=3.22\times10^{23}$ W\,Hz$^{-1}$. The average surface brightness is $\sim$0.09~$\mu$Jy/arcsec$^{2}$.

There is strong interest in finding bridges between clusters; they are
often not detected, despite high sensitivity \citep[i.e. see the recent ASKAP search of the pre-merger cluster pair A\,3391-A\,3395;][]{Bruggen21}.
Successful detections of bridges between cluster pairs have been made using LOFAR at 144~MHz, such as the cluster pair A\,1758N--A\,1758S \citep{Botteon20a} and the system A\,399--A\,401 \citep{Govoni19}. In addition, there is a well-known bridge connecting the Coma cluster with the NGC\,4839 group, detected at 144 MHz \citep{Bonafede21} and at 346 MHz \citep{Kim89}.
All these clusters are considerably more massive than A\,3562 and SC\,1329--313,
and the detections are well below the GHz frequencies.

\subsection{Radio source J~1332--3146a}\label{sec:butterfly}

\begin{figure*}[h!]
\centering
{\includegraphics[angle=-90,scale=0.3]{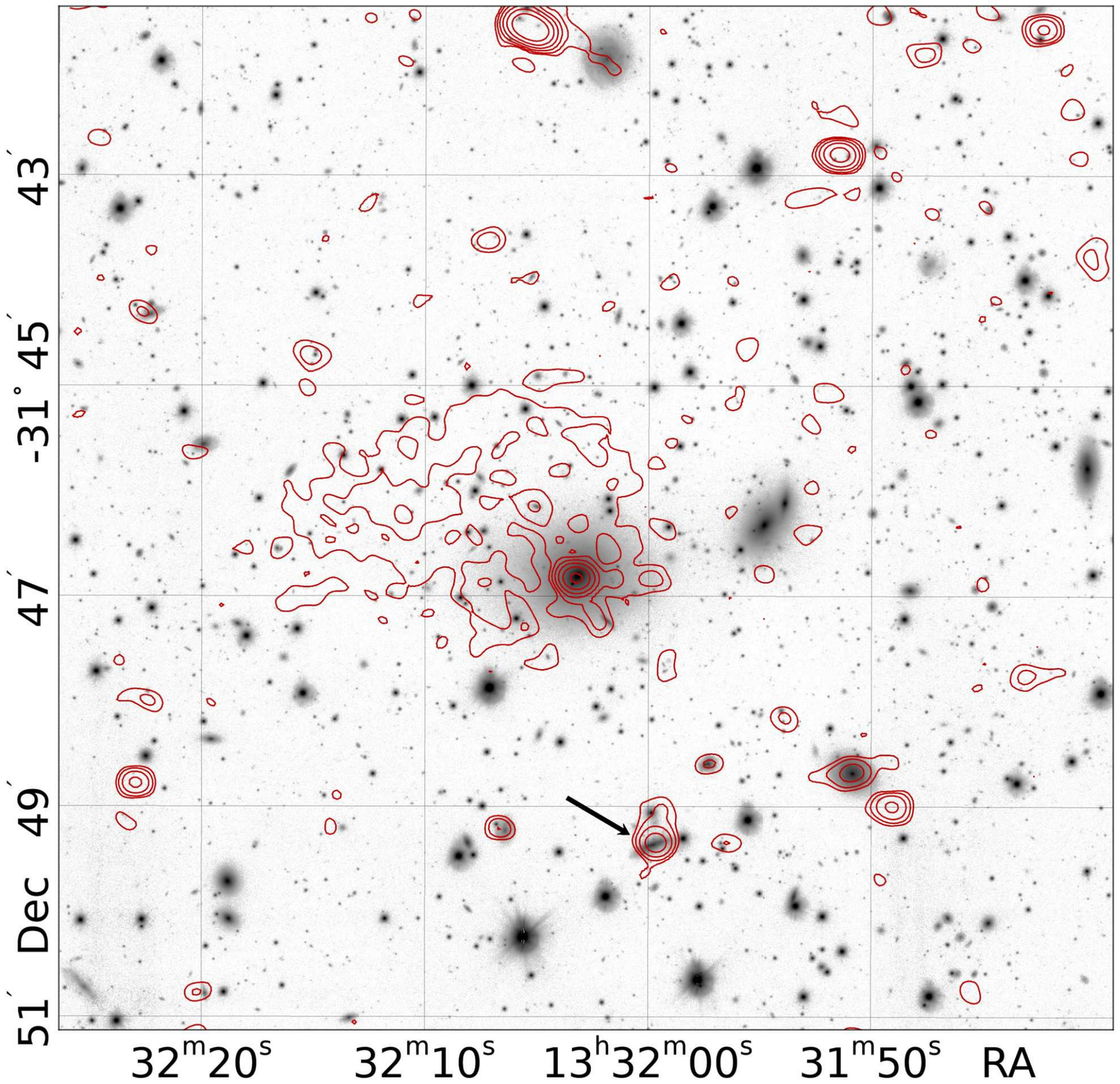}
\hskip 0.1truecm
\includegraphics[angle=-90,scale=0.3]{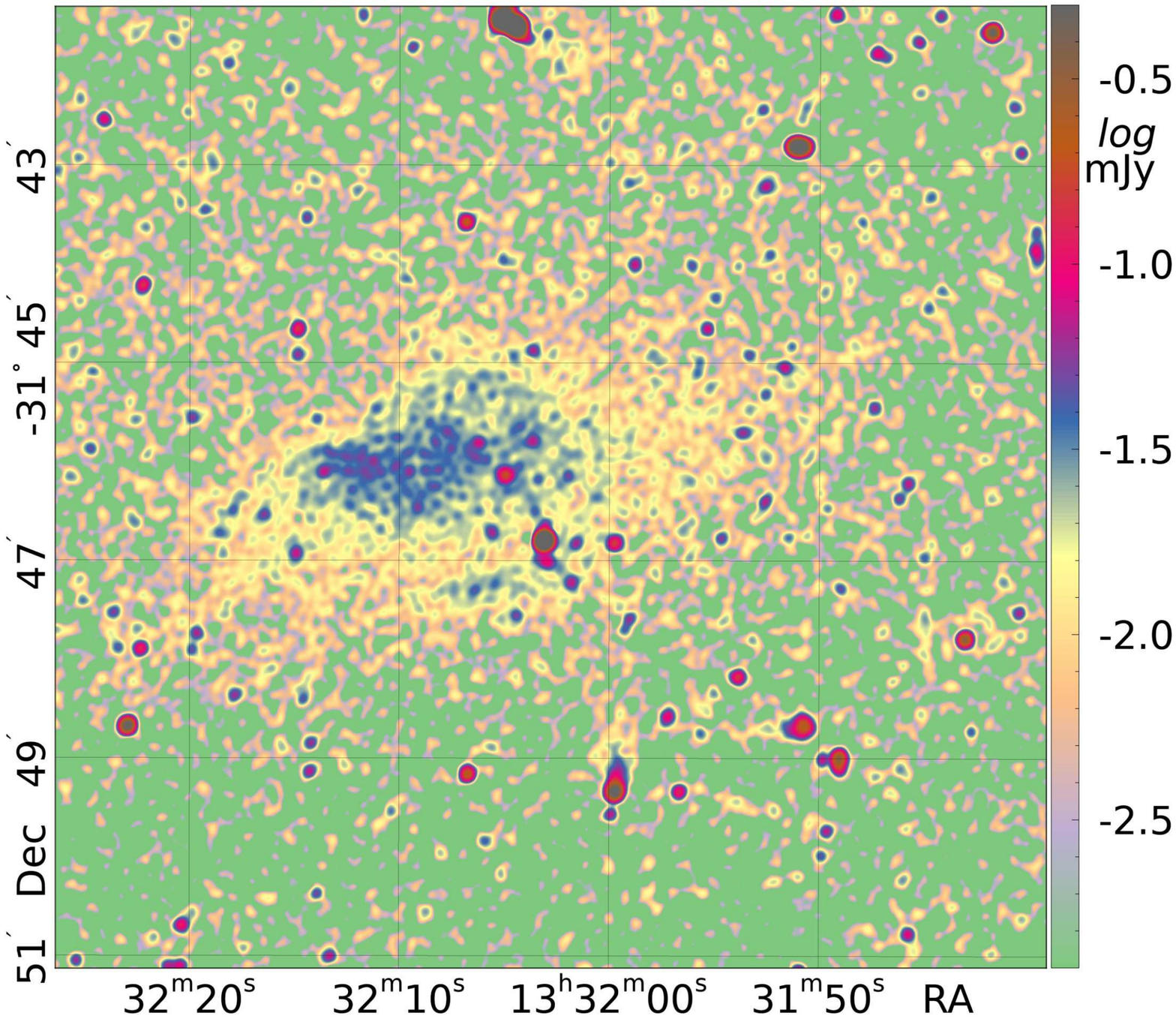}}
\caption{Contours of the 887\,MHz ASKAP image zoomed around
  J\,1332--3146a superimposed on the VST $r$-band image (left panel). The restoring beam is
  $13.2^{\prime\prime} \times 10.4^{\prime\prime}$, p.a. $85.4^\circ$. The first contour is
  drawn at 0.1 mJy~beam$^{-1}$, and the contours are spaced by a factor of 2.
  SOS\,61086, located south of J\,1332-3146a, is indicated by the black arrow.
  The first contour is drawn at $\sim$3$\sigma$. Colour scale of the 1.283\,GHz
  MeerKAT image (same as central and bottom panel of Fig. \ref{fig:fig3}),  reported
  to emphasise the structure of the radio emission (right panel).}
\label{fig:fig7}
\end{figure*}

The radio source J\,1332--3146a was first imaged with the VLA at 1.4 GHz (G05).
The source is located just east of the X--ray emission of SC\,1329--313 (G05 and Fig. \ref{fig:fig5}).
The new ASKAP and MeerKAT images presented here are in very good agreement with the previous ones, but the much better sensitivity and resolution of  the current observations provide new insights on the morphology and surface brightness distribution of the emission. The overall size of the source, 6$^{\prime}\times2.5^{\prime}$ ($\sim$ 335 $\times$ 140 kpc), confirms the previous measurement.
Figs. \ref{fig:fig3} and \ref{fig:fig7} clearly show that no jets or other features link the diffuse emission to the compact component, at least at the resolution of our images. We further see no evidence for a connection between the embedded strong compact radio source and the extended emission in the spectral index distribution.
The morphology of J\,1332--3146a is not symmetric  with respect to the compact component but extends north and east, with a 1283 MHz (MeerKAT) surface brightness in the range $\sim$0.14 - 0.75\,$\mu$Jy/arcsec$^2$.

The embedded bright compact source is located at the southwest end of the radio emission (Fig. \ref{fig:fig7}, left panel) and is associated with a bright early-type galaxy of magnitude\footnote{Corrected for Galactic extinction.} $r=13.56$ with redshift z=0.04351, belonging to SC\,1329--313 \citep{Haines18}.
The galaxy (RA$_{\rm J2000}= 13^h32^m03.17^s$, DEC$_{\rm J2000}= -31^{\circ}46^{\prime}48.5^{\prime\prime}$) is located along the axis connecting the peaks of the X--ray emission in A\,3562 and SC\,1329--313 (see Fig. \ref{fig:fig5} and Fig. 9 in G05).

\begin{figure*}[h!]
{\includegraphics[scale=0.43]{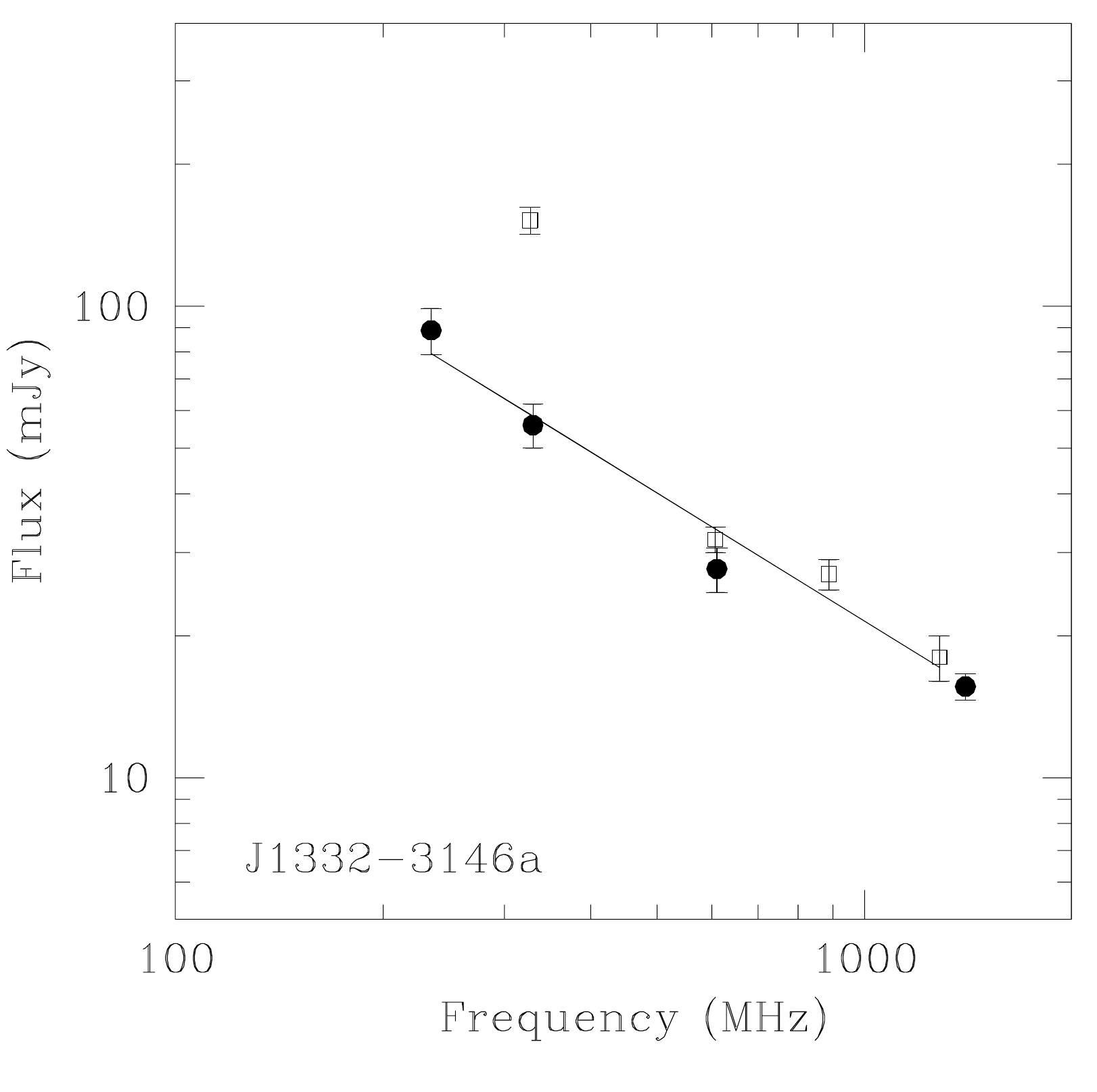}
\hskip 0.3truecm
\includegraphics[scale=0.35]{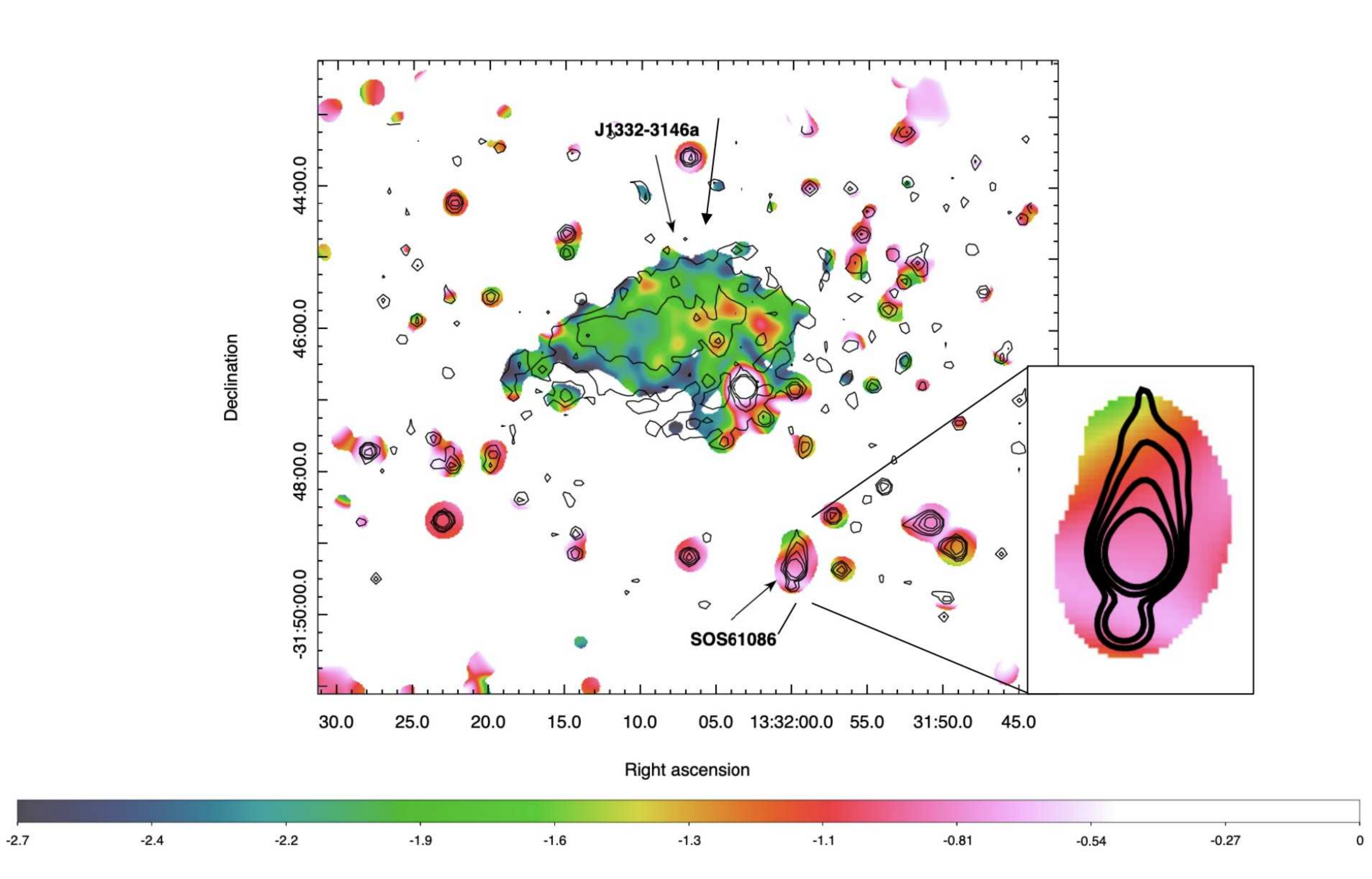}}
\caption{Radio spectrum of the radio source J1332--3146a, obtained using the data in
  Table \ref{tab:fluxbutterfly} (left panel). Filled dots are from G05, open squares are from the present paper.
  The fit does not include the 306 MHz measurement (see Sect. 4.3).
  In-band MeerKAT spectral image at the resolution of $15^{\prime\prime}\times15^{\prime\prime}$ (right panel).
  The radio contours are from the full resolution image, i.e. $7.3^{\prime\prime}\times7.2^{\prime\prime}$,
  and are drawn at $\pm$20, 40, 80, and 160\,$\mu$Jy\,beam$^{-1}$.
  A zoom of the radio source associated with SOS\,61086 is shown in the inset.}
\label{fig:fig8}
\end{figure*}

We used all the datasets presented here to derive the integrated spectrum, and complemented them with the information published in G05.
Table \ref{tab:fluxbutterfly} reports the total flux density measurements (after removal of the compact emission associated with the galaxy) and the spectrum is shown in the left panel of Fig. \ref{fig:fig8}.
The flux density measurements of J\,1332--3146a are considerably scattered.
The sensitivity of the GMRT 233 MHz observations presented in this paper is too low to detect this source. The 306 MHz flux density value does not align with the data points at higher frequencies and it is not consistent with our previous measurement either. The reason for this discrepancy is unclear.
Since the radio source is located at the very edge of the GMRT field of view at this frequency, the local noise is higher and as a consequence the uncertainty on the total flux density is larger than that of the other diffuse sources presented here.
If we ignore this measurement, a linear fit to the data provides a value $\alpha_{\rm 235~MHz}^{\rm 1283~MHz}=-0.76\pm0.2$. A considerable steepening above 1 GHz is suggested by the MeerKAT in-band spectral index image shown in the right panel of Fig. \ref{fig:fig8} (derived from the
15$^{\prime\prime}\times15^{\prime\prime}$ resolutions images of pointing (d) in SC\,1329--313, see Table 2), which covers the frequency range 908-1656 MHz.
\\
The distribution of the in-band spectral index{\bf \footnote{To fit the in-band spectral indices, we used a new spectral fitting code (part of the PFB-Clean (https://github.com/ratt-ru/pfb-clean suite).
This proceeds as follows. The per-subband (8, in this case) model and residual images produced by wsclean are reconvolved to a common resolution. A power law is then fitted pixel-by-pixel (weighting the subbands as per the ``wsum`` FITS keyword generated by wsclean). The subband images are generated by the same run of wsclean as the MFS image, and use a weighting of robust=0 (without the ``use MFS weighting'' option of wsclean in effect, so the weighting is truly robust=0 per subband). The primary beam is accounted for by directly incorporating attenuation by the average Stokes I beam into spectral index model during the fitting. We used the Eidos beam model (https://github.com/ratt-ru/eidos \cite{Asad21} in this particular instance).}}
strengthens the idea that the compact radio source and the extended emission are not connected. The former is flat ($\alpha \sim$\,--0.3 in Fig. \ref{fig:fig8} ), while the diffuse emission is considerably steeper, with $\alpha$ in the range [--2.3,--1] across the source.
Comparison of the right panel of Figs. \ref{fig:fig6} and \ref{fig:fig7} further shows that the region of the diffuse emission where the spectrum is flatter (pink) are those with embedded compact sources, whose spectrum is less steep than the rest. The possible origin of this emission is discussed in Sect. 5.1.1.

%
%

%
%

%
%
%
%
\begin{table}[h!]
  \caption[]{Flux density values of J1332-3146a.}
\begin{center}
\begin{tabular}{lclc}
\hline\noalign{\smallskip}  
$\nu$ & Array & Flux & Ref \\
MHz   &       & mJy  &     \\
\hline\noalign{\smallskip}
1400 & VLA     &  16$\pm$1 & G05 \\
1283 & MeerKAT &  18$\pm$ 2 & This work \\
887  & ASKAP   &  28$\pm$ 2 & This work \\
610  & GMRT    &  28$\pm$3 & G05 \\
607{\bf $^{\rm (a)}$}  & GMRT    &  32$\pm$ 2 & This work \\
{\bf 306$^{\rm (b)}$}  & GMRT    & 152$\pm$39 & This work \\
330  & GMRT    &  56$\pm$6 & G05 \\
235  & GMRT    &  89$\pm$10 & G05 \\ 
\hline\noalign{\smallskip}
\end{tabular}
\end{center}
    {Notes: $^{\rm (a)}$ Project 30\_024 and $^{\rm (b)}$ Project 22\_039
        in Table \ref{tab:logs}.}
\label{tab:fluxbutterfly}
\end{table}
%
%
%
%

\subsection{Radio tail of SOS~61086}

We identify a tailed radio source south-west of J\,1332--3146a (Fig.~\ref{fig:fig7}, left panel, and Fig. \ref{fig:fig8}, right panel), associated with a spiral galaxy member of SC\,1329--313\footnote{The membership has been statistically assigned by means of the dynamical analysis in \cite{Haines18}.} identified as SOS\,61086 in the Shapley Optical Survey
\citep{Mercurio06,Haines06}.
The galaxy has a redshift $z=0.04261\pm0.00023$ ($V_{h}=12500\pm 70$\,km\,s$^{-1}$) and is located at $\sim$280\,kpc ($\sim 0.3 r_{200}$) in projection from the X-ray centre.
Considering the median redshift of SC\,1329--313, the line-of-sight peculiar velocity of SOS\,61086 with respect to the main group is $-900$\,km\,s$^{-1}$.
The radio source was detected at 1.4 GHz in \cite{Giacintucci04} and labelled J\,1331--3149b; it was classified as unresolved at the resolution and sensitivity of those observations.

\begin{figure*}[h!]
\centering
{\includegraphics[scale=0.33]{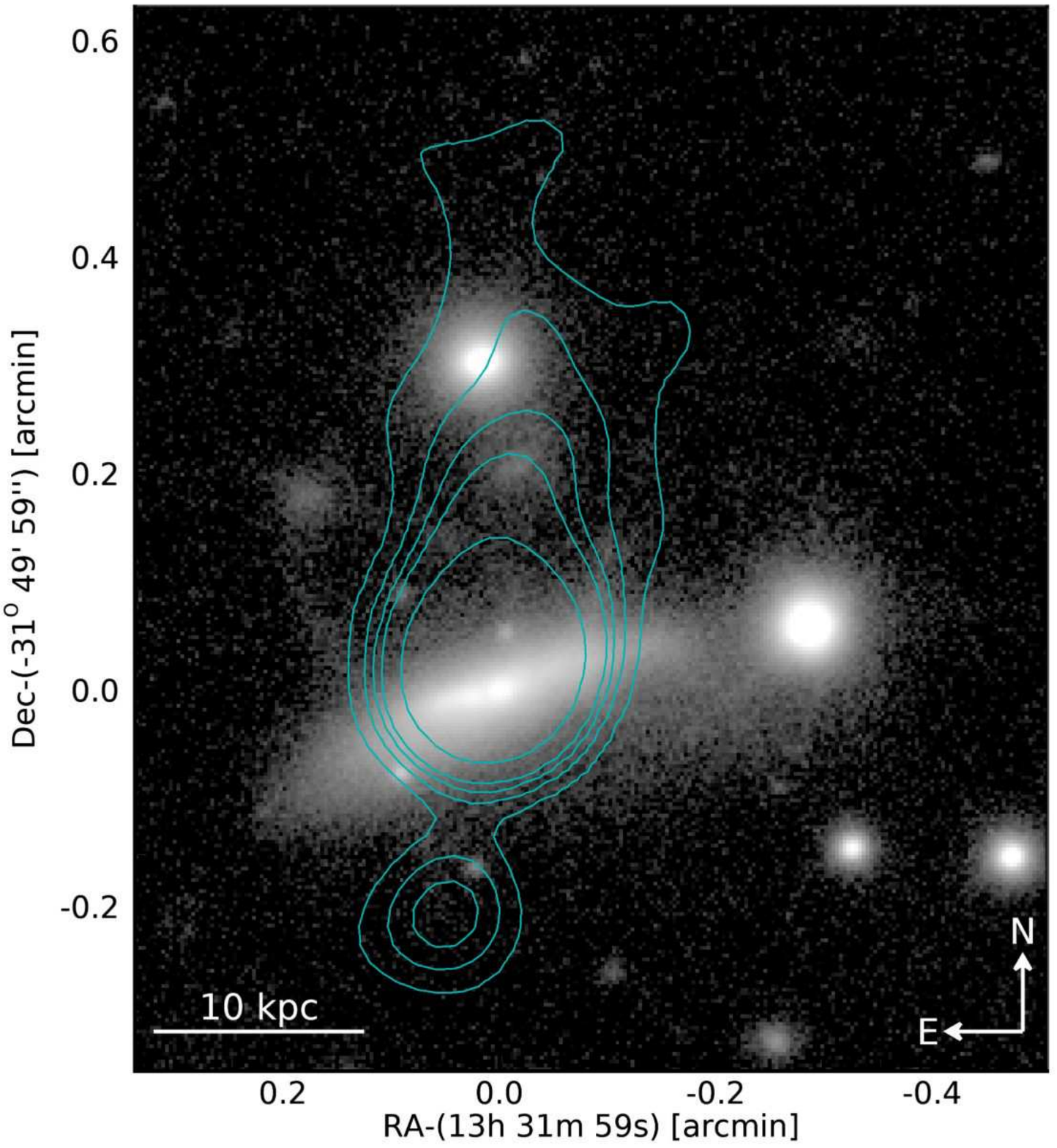}
\hskip 0.5truecm
\includegraphics[scale=0.5]{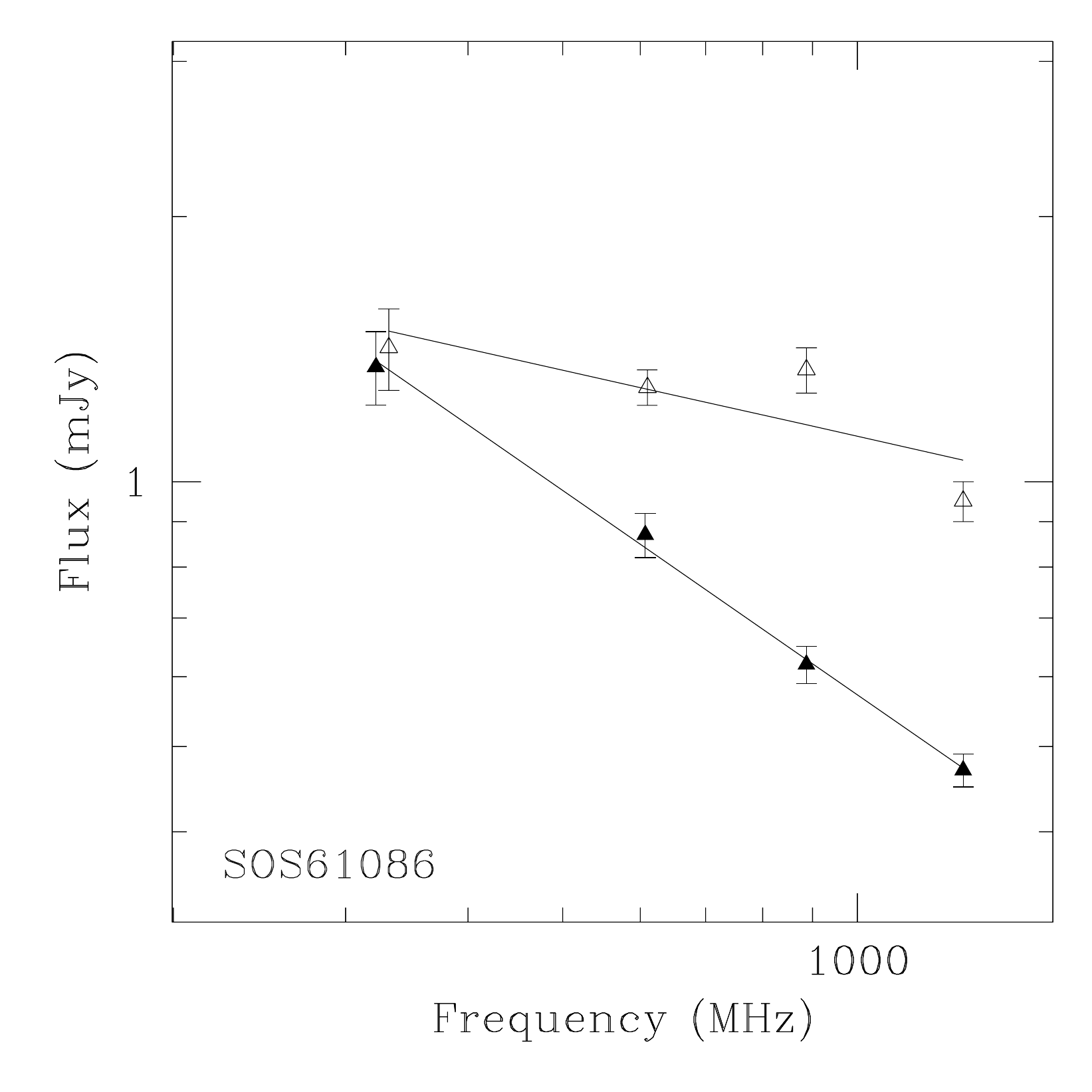}}
\caption{ESO-VST  $r$-band image of the field including
  SOS\,61086 with 1.28 GHz MeerKAT
  radio contours superposed (left panel). The restoring beam of the radio image is
  $6.9^{\prime\prime} \times 6.5^{\prime\prime}$, p.a. $151.8^\circ$. The contours are plotted at 20, 40, 60,
  160\,$\mu$J~beam$^{-1}$. The first contour is drawn at $\sim$\,3.5$\sigma$ (the local noise is
  $\sim$\,6\,$\mu$Jy\,beam$^{-1}$).
  The scales (RA and DEC) give the distance (in arcsec) from the photometric
  centre of the galaxy. Spectrum of the radio emission of SOS\,61086 (right panel). Open and
  filled triangles are the emission from the galaxy and the tail, respectively.}
\label{fig:fig9}
\end{figure*}

The $r$-band image of the galaxy from the ShaSS \citep{Merluzzi15} is shown in Fig.~\ref{fig:fig9} (left panel) superimposed with the contours of the 1.28\,GHz MeerKAT image. Because of the hints of matter beyond the galaxy stellar disk in the northerly direction, the object has been investigated in detail with observations carried out with the Wide-Field Spectrograph \citep{Dopita07,Dopita10} mounted at the Nasmyth focus of the Australian National University 2.3m telescope located at Siding Spring Observatory (Australia), complemented by N-body/hydrodynamical simulations \citep[for the full analysis see][]{Merluzzi16}. This study demonstrated that the galaxy of stellar mass $\mathcal{M}_ {\star}\sim4\times10^{9}$M$_\odot$ is undergoing ram-pressure stripping (RPS) and suggested that the time of the onset (`age') of RPS is about 250\,Myr ago. Furthermore, the onset epoch estimate agrees with the age of the young stellar population ($< 200$\,Myr), suggesting that we are very likely observing
ram-pressure-induced star formation.

The new high-quality and deep MeerKAT observations reveal the radio tail, shown in the left panel of Fig.~\ref{fig:fig9}. The radio continuum emission peaks at the galaxy centre, and extends north at comparable brightness levels up to 30\,kpc in projection from the galaxy disk.
The other striking feature of the radio continuum emission is its confinement to the inner part of the disk which recalls the truncation of the gas disk in galaxies affected by RPS.

We used our images to derive the spectrum of the radio emission associated with the galaxy disk and that of the tail, which are shown in the right panel of Fig.~\ref{fig:fig9}. The spectrum of the tail (computed using  images at similar angular resolution) is considerably steeper than the emission from the galaxy and it is very well fit by a power law with $\alpha_{\rm 306~MHz}^{\rm 1283~MHz}=-0.79\pm0.05$. The flux density values of the disk are more scattered, and this could be partly due to the different resolutions of the images. We further note that the MeerKAT in-band spectral index of SOS\,61086, shown in the inset of the right panel of Fig.~\ref{fig:fig8}, clearly shows a steepening along the tail, up to $\alpha\sim$\,--2.
Assuming that the spectrum of the tail steepens above $\sim$1.2 GHz, as suggested from the in-band spectral index, we derive a magnetic field H$_{\rm eq}$=0.9\,$\mu$G, and an upper limit to the radiative age of the relativistic electrons of $\sim$ 100 Myr (including the aging of the electrons due to scattering with the CMB).
Finally, Figs. \ref{fig:fig5} and \ref{fig:fig7} (right panel) suggest that a much fainter diffuse intercluster radio emission extends north  of SOS\,61086 to J1332--3146a, over an extent of $\sim$ 90 kpc.

%
%

\subsection{Head--tail in SC~1329--313}

The ASKAP and MeerKAT images in Fig. \ref{fig:fig3} show an intriguing emission west of J1332--3146a, at RA$\sim 13^h30^m40^s$.
A zoom into that area is given in Fig. \ref{fig:fig10}.

The head--tail radio source is associated with a cluster galaxy located at RA$_{\rm J2000}=13^h30^m48.91^s$, DEC$_{\rm J2000}=31^{\circ}43^{\prime}25.6^{\prime\prime}$, with z=0.044 and magnitude R=13.6 (from the NASA/IPAC Extragalactic Database). However, a superimposed foreground star makes the magnitude measurement unreliable. The galaxy and the star can be distinguished
only in the VISTA $K$-band image of ShaSS.
The two jets of the tail bend sharply just outside the envelope of the optical galaxy, and lose their collimation at a distance of $\sim$40 kpc from the core, to form a feature of very low surface brightness emission, which we refer to as the broken tail, extending $\sim$250 kpc southwards. At that location the morphology of the emission changes again, to take the shape of a bar inclined at $\sim 30^{\circ}$  with respect to the orientation of the broken tail. The length of the bar is $\sim$250 kpc.
Part of this emission was detected with the GMRT at 306 MHz (V17), however, the sensitivity of those observations did not allow for a clear morphological classification of the source.

Table \ref{tab:fluxheadtail} reports the flux density of all these components in our datasets. At 1283 MHz, we considered the values derived from the MGCLS (pointing (d) in Table \ref{tab:logs}) to avoid any possible uncertainties in the primary beam correction at very large distance from the pointing centre. At 607 MHz we reported the values at different resolutions to ensure the full detection of the extended features.
Only the inner part of the tail and the bar are detected at 233 MHz. One possibility is that the core is self-absorbed at frequencies below 306 MHz, while the broken tail and the southern emission most likely fall below the sensitivity limit of those observations.
All the remaining features are clearly detected in all datasets.

The spectra of the various features are shown in the left and right panels of Fig. \ref{fig:fig11}.
The spectrum shown in the left panel includes both the core and the inner part of the head-tail, corresponding to the red area in the left panel of Fig. \ref{fig:fig10}. The irregular trend is most likely the result of the different components in the core and jets, which are difficult to disentangle.
Overall this region shows an approximately flat spectrum.
The situation is completely different in the broken tail and in the bar.
Both features are steep, and a linear fit provides $\alpha_{\rm 306~MHz}^{\rm 1283~MHz}=-1.0\pm0.1$ in the broken tail, and $\alpha_{\rm 306~MHz}^{\rm 1283~MHz}=-1.95\pm0.05$.

The spectral index values of the various features are confirmed by the MeerKAT spectral index image shown in the left panel of Fig. \ref{fig:fig12} (derived from pointing (b) in SC\,1329--313, see Table 2). The core of the head--tail has a spectral index $\alpha\sim -0.5$, with a clear steepening along the jets, and the remarkable steepness of the bar is confirmed, with $\alpha$\simlt --2. 

We used the pointing from the MGCLS (see Table \ref{tab:logs}) to search for polarisation information on the head-tail and the bar. Preliminary Q and U images from the MGCLS were constructed by summing the respective individual frequency channels and, thus, they do not correct for any Faraday rotation, spectral index, or depolarisation effects. The approximate fractional polarisations are shown in the right panel of Fig. \ref{fig:fig12}.
With the above caveats, the observed fractional polarisation is most prominent in the eastern tail, and increases from  of order 10\% near the core to an approximate level of 70\%, which is the theoretical maximum, for the bar.
More precise results would require a full Faraday synthesis, which is beyond the scope of the current work.

\subsection{Southern emission}

\begin{figure*}[h!]
\centering
{\includegraphics[scale=0.235]{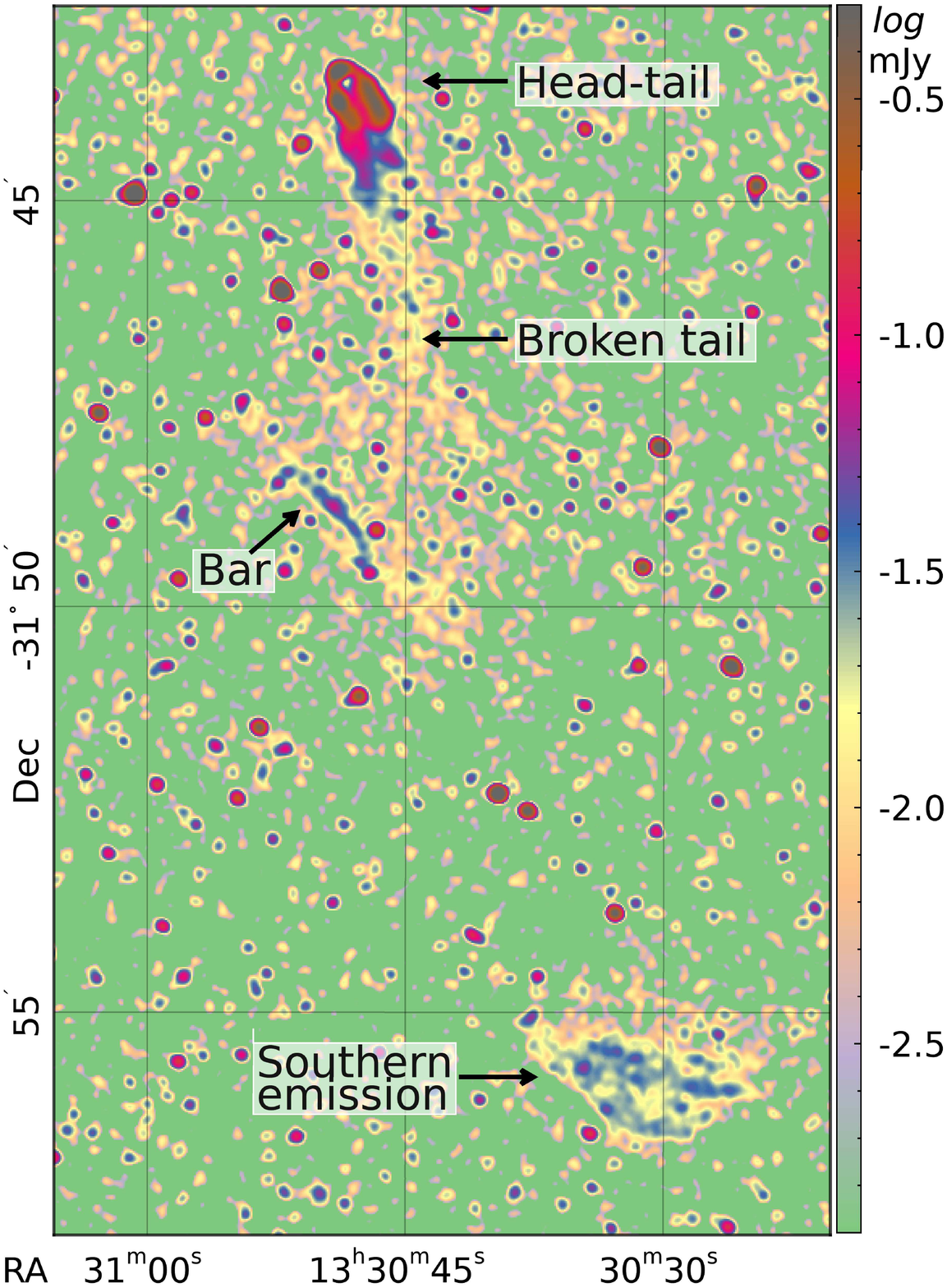}
\hspace{2.5mm}
       \includegraphics[scale=0.23]{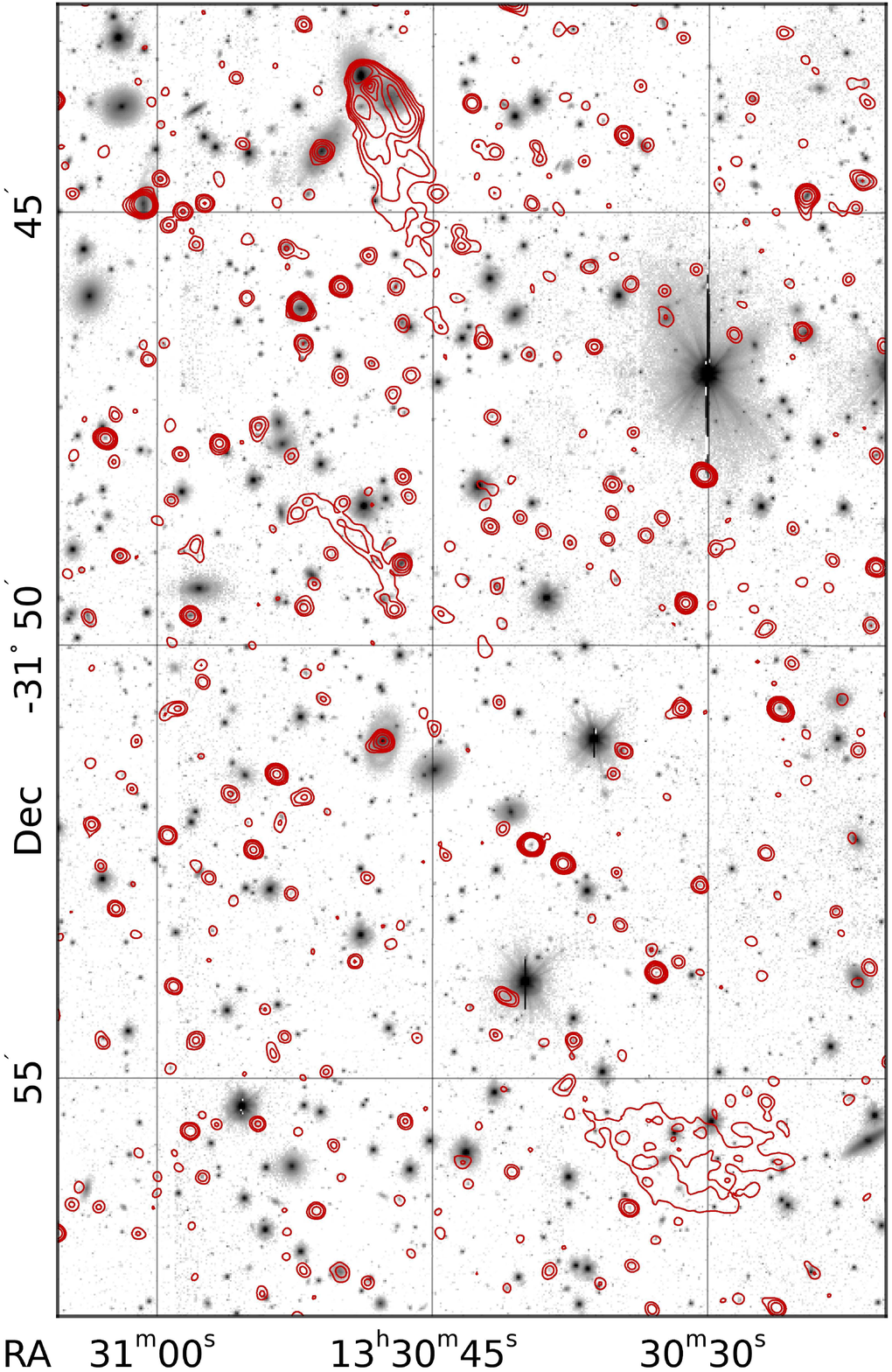}
       \hspace{2.5mm}
              \includegraphics[scale=0.235]{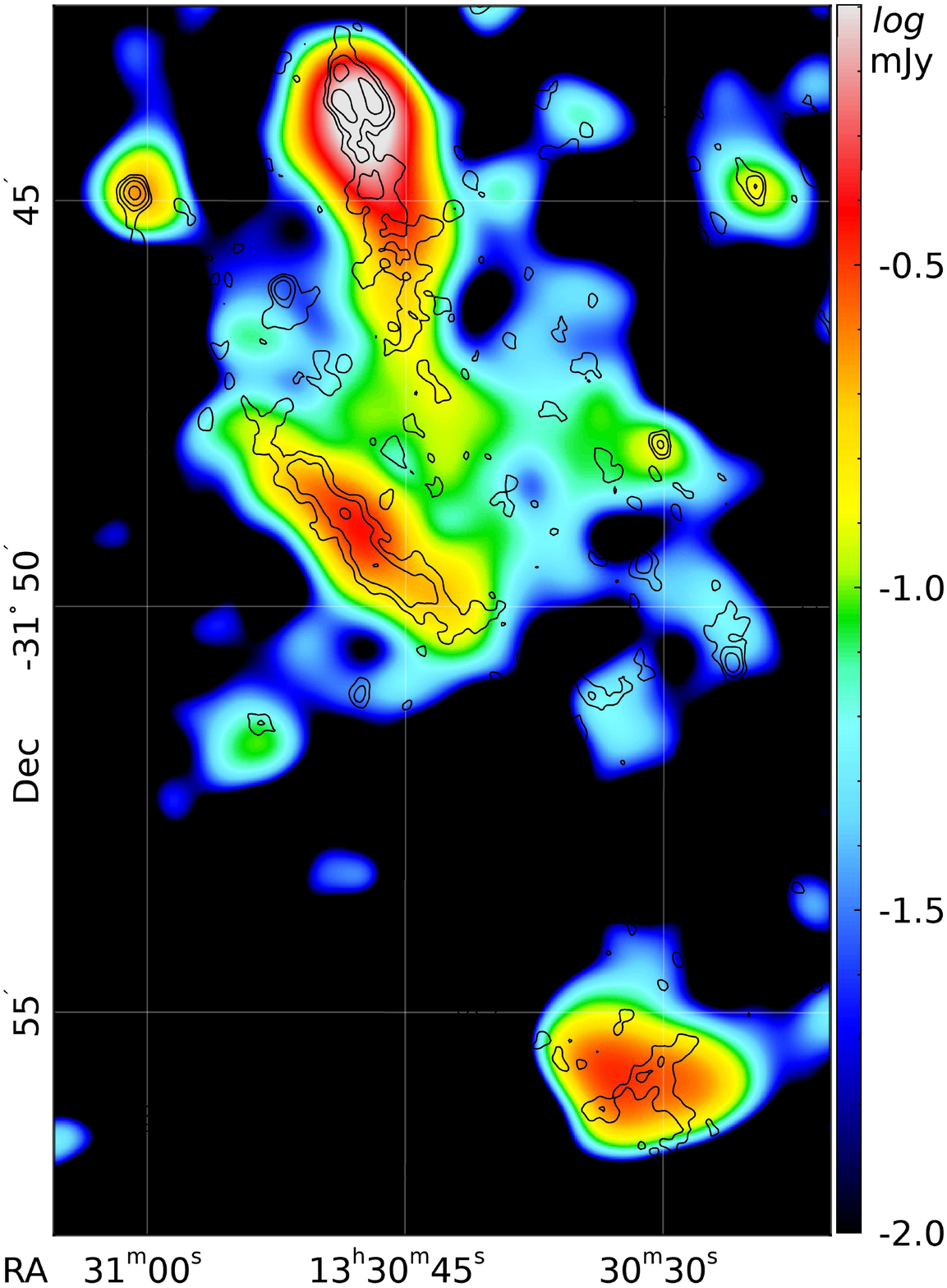}}
\caption{1.283 GHz MeerKAT image of the head-tail in SC\,J1329--313 (left panel).
  The colour scale is in the range -1 - 0.1 mJy~beam$^{-1}$.
  The restoring beam of the image is $7.66^{\prime\prime}\times7.28^{\prime\prime}$,
  in p.a. $68^{\circ}$. The various features of diffuse emission are indicated.
   Contours of the same image overlaid on the ESO-VST $r$-band image (central panel). The
  contour levels are $\pm$20, 40, 80, 160\,$\mu$Jy~beam$^{-1}$. The first contour
  corresponds to $\sim 3\sigma$. Right panel - 1.283 GHz MeerKAT image of the diffuse
  emission at the resolution of $40^{\prime\prime}\times40^{\prime\prime}$ in colour with
  GMRT contours at 306 MHz overlaid. The contours are drawn at 0.2, 0.4, 0.8 mJy\,beam$^{-1}$.
  The angular resolution is $14.0^{\prime\prime}\times9.5^{\prime\prime}$.}
  \label{fig:fig10}
\end{figure*}

\begin{figure*}[h!]
\centering
{\includegraphics[scale=0.5]{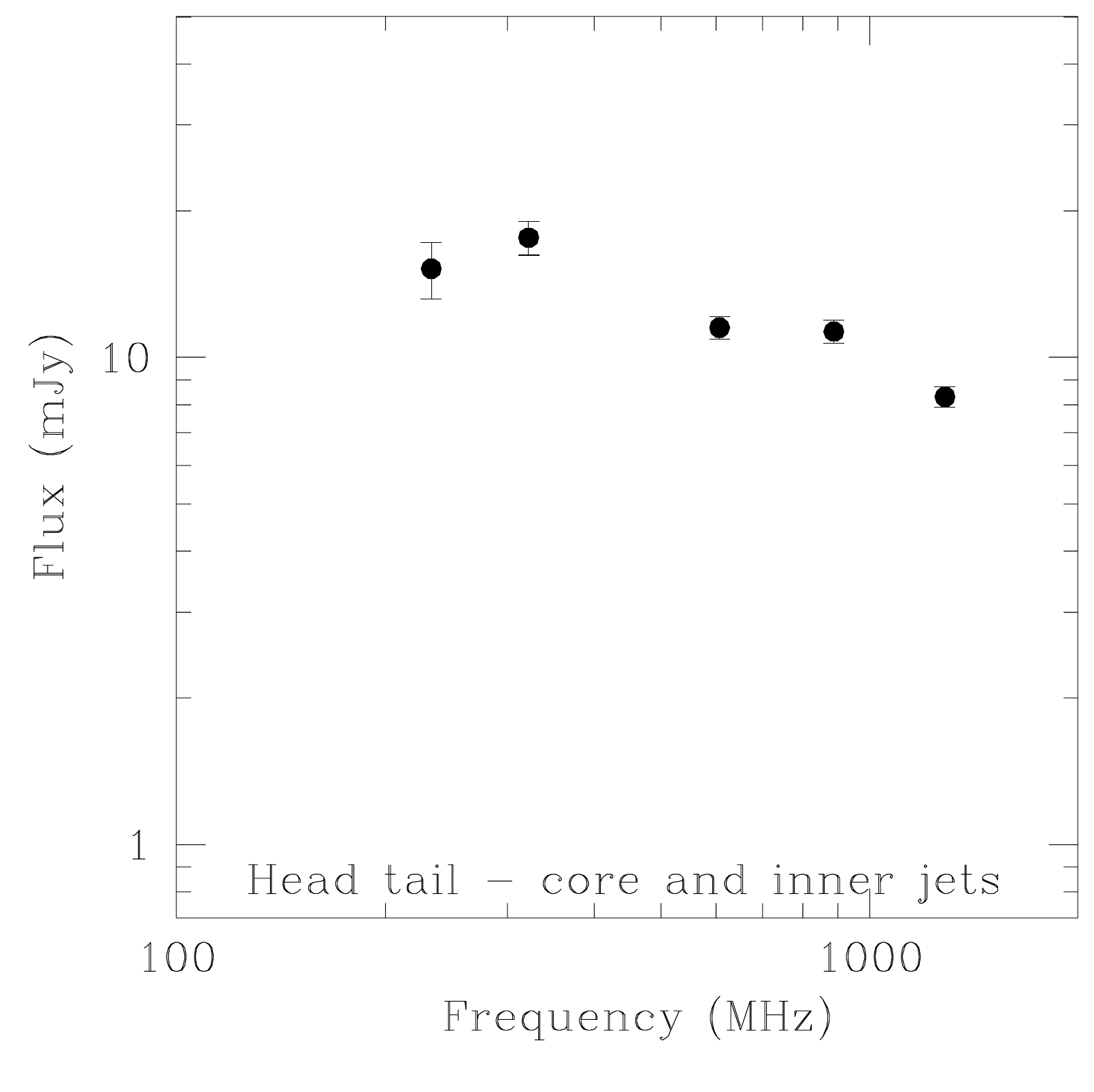}}
\hspace{6mm}
{\includegraphics[scale=0.5]{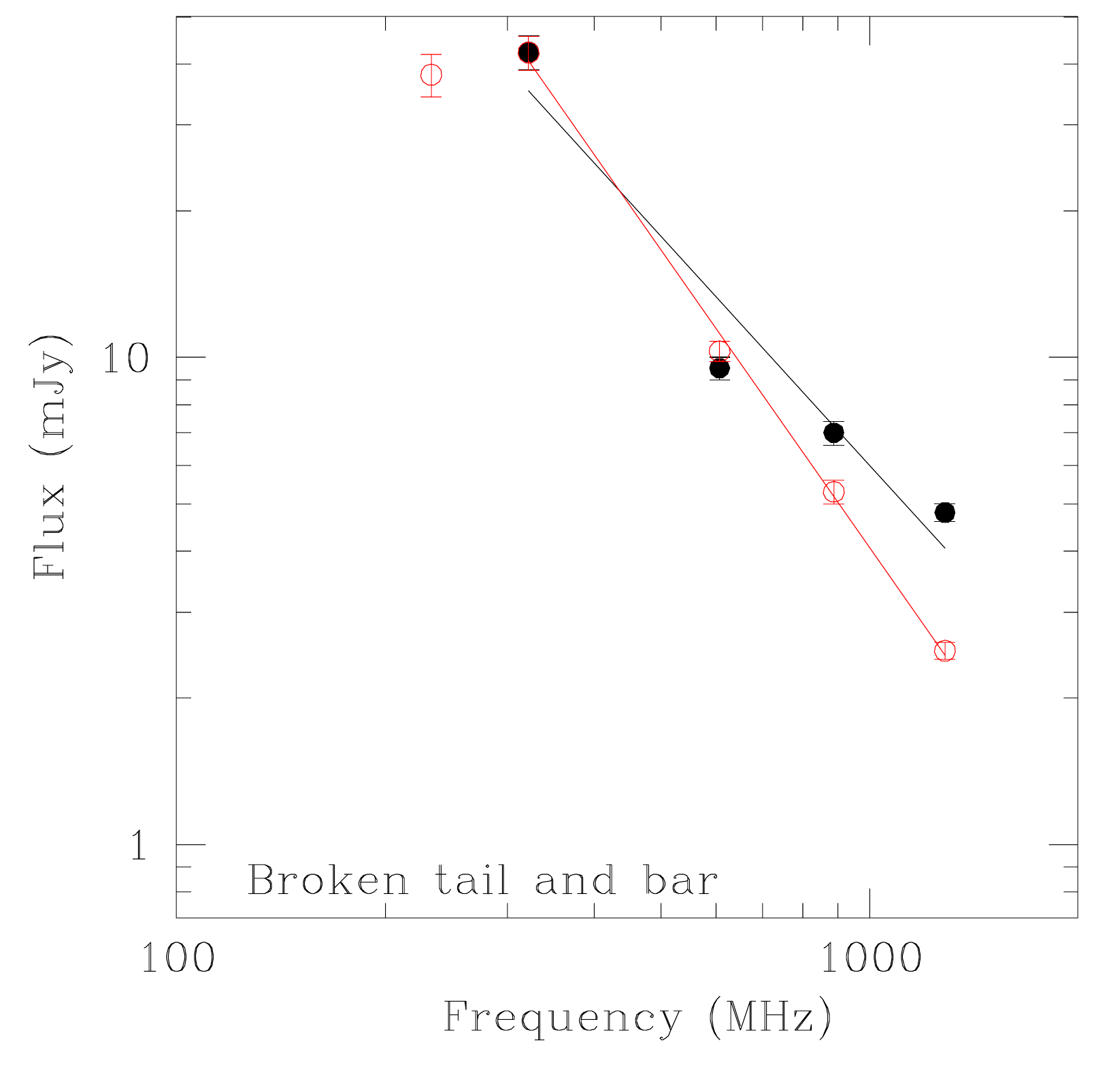}}
\caption{Spectra of the components in the head--tail radio galaxy.
  Left panel:\ Core region, encompassing the core and the inner tail (red area in the left
  panel of Fig. \ref{fig:fig10}). Right panel:\ Broken tail (black filled circles) and bar
  (red open circles).}
  \label{fig:fig11}
\end{figure*}

\begin{figure}[h!]
\includegraphics[scale=0.35]{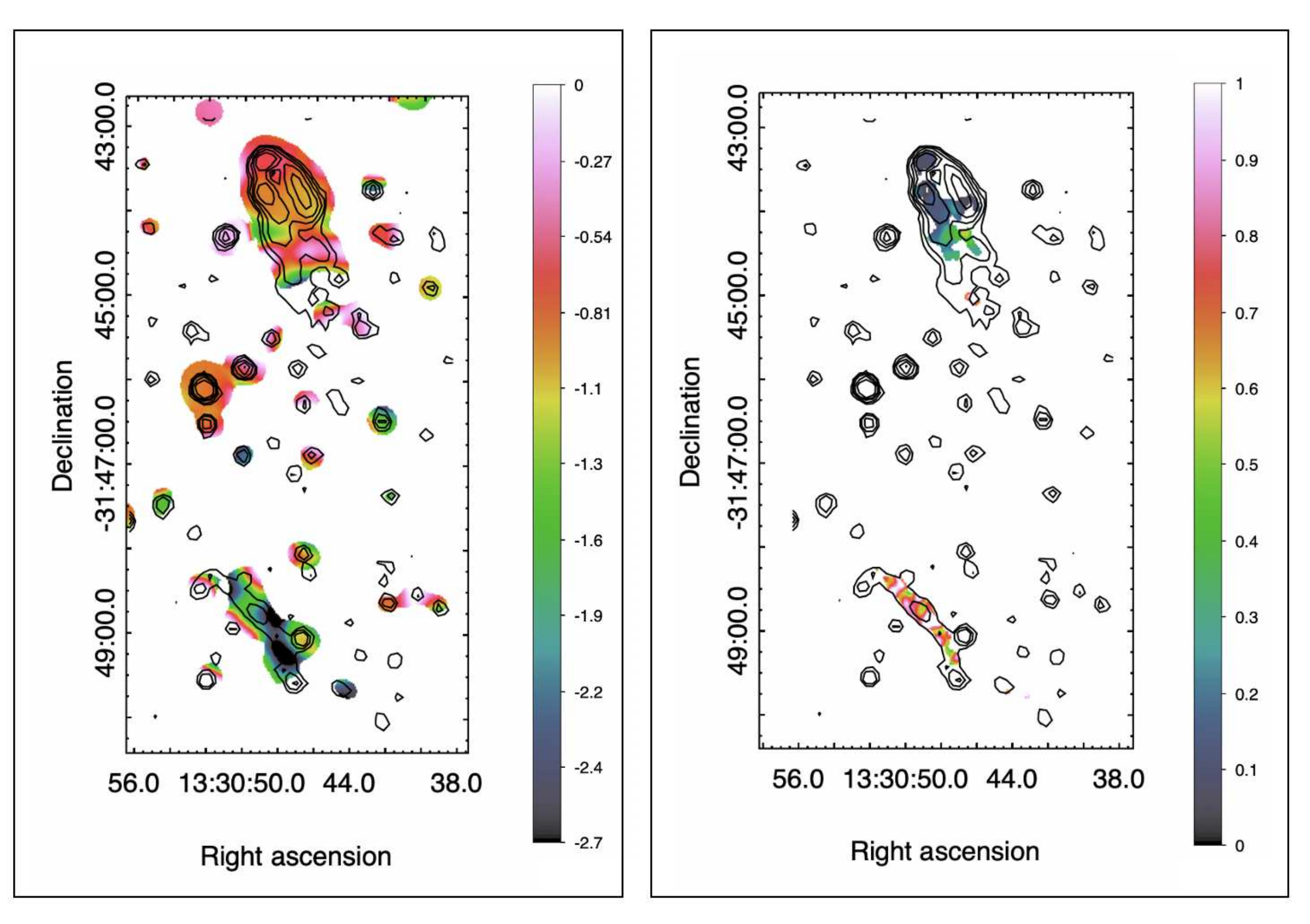}
  \caption{MeerKAT 1-1.8 GHz in-band spectral image of the head-tail, broken
    tail, and bar at the resolution of 15$^{\prime\prime}\times15^{\prime\prime}$ (left panel).
  Contours are the same as in the central panel of Fig. \ref{fig:fig10}.
  In the right panel we show the same contours as in the left panel overlaid on the fractional
  polarisation shown in colour.}
  \label{fig:fig12}
\end{figure}

This interesting feature is not obviously connected to any optical source, as is clear from the central panel of Fig. \ref{fig:fig10}.
The surface brightness is quite uniform  and its spectrum, with $\alpha_{\rm 607~MHz}^{\rm 1283~MHz}= -0.35\pm0.12$, is unusually flat for diffuse sources with no optical counterpart.
Fig. \ref{fig:fig5} shows that it is outside the region covered by the {\it XMM-Newton} observations.

These properties challenge its classification within the current classes of extragalactic radio sources.
Assuming that the source is located at the distance of the Shapley Supercluster, its size is $\sim167\times83$ kpc.
The source is too faint to allow in-band spectral imaging.

We note that the brightest optical/infrared object co-located with the southern emission is a giant star 
\citep[WISEA J133029.79, GAIA 6169476948,][]{Stassun19} with a high proper motion ($-4.72\pm0.05$ mas year$^{-1}$ in RA and $-2.41\pm0.04$ mas year$^{-1}$ in Dec).
At a distance of 3850$\pm$400 pc, this corresponds to a velocity of $\sim$100 km s$^{-1}$.
If the flat-spectrum southern emission were associated with this star, it would have a size of $\sim$4$\times$2.5~pc, and a 1~GHz monochromatic luminosity of 6.4$\times 10^{12}$ W\,Hz$^{-1}$. For comparison, this is 40 times the surface area and 10 times the luminosity of the optically thin free-free emission from the outer portion of the mass loss wind from \emph{P Cygni} \citep{Skinner98}.

%
%

%
%
%
%


\begin{table*}[h!]
  \caption[]{Flux density values of the head-tail and southern feature.}
\begin{center}
\begin{tabular}{ccccccc}
\hline\noalign{\smallskip}  
$\nu$ & Array & Resolution & Core+ Inner tail & Broken tail & Bar & Southern emission \\
MHz   &       & $^{\prime\prime}\times^{\prime\prime}$ & mJy & mJy & mJy & mJy  \\
\hline\noalign{\smallskip}
1283 & MeerKAT &  7.66$\times$7.26 & 8.3$\pm$0.4 & 4.8$\pm$0.2 &  2.5$\pm$0.1 & 3.7$\pm$0.1 \\
887  & ASKAP   & 13.23$\times$10.43&  11.3$\pm$0.6 & 7.0$\pm$0.4 &  5.3$\pm$0.3 & 4.0$\pm$0.2 \\
607{\bf $^{\rm (a)}$}  & GMRT    &  9.73$\times$5.55 &  11.5$\pm$0.6 & 9.5$\pm$0.5 & 10.3$\pm$0.5 & -           \\
607{\bf $^{\rm (a)}$}  & GMRT    &  15$\times$15     &  12.8$\pm$0.6 &10.9$\pm$0.5 & 10.7$\pm$0.5 & 4.8$\pm$0.2 \\
{\bf 306$^{\rm (b)}$}  & GMRT    & 14.05$\times$9.53 &  17.6$\pm$1.4 &42.3$\pm$3.4 & 42.2$\pm$3.4 & -           \\
233  & GMRT    & 24.45$\times$10.77&  15.2$\pm$1.5 &  -          & 38.0$\pm$3.8 & -           \\
\hline\noalign{\smallskip}
\end{tabular}
\end{center}
{Notes: $^{\rm (a)}$ Project 30\_024 and $^{\rm (b)}$ Project 22\_039
        in Table \ref{tab:logs}.}
\label{tab:fluxheadtail}
\end{table*}
%
%
%

\subsection{Diffuse emission at the centre of A~3558}

A\,3558 is located $\sim$2.8 Mpc west of SC\,1329--313 and  
is the most massive cluster in the Shapley Supercluster
(Table \ref{tab:info}). Its size and overall dynamical state as derived from the X--ray properties \citep{Rossetti07} make it an ideal candidate to host diffuse emission in the form of a radio halo, undetected prior to our observations.

\begin{figure}[h!]
\centering
{\includegraphics[scale=0.48]{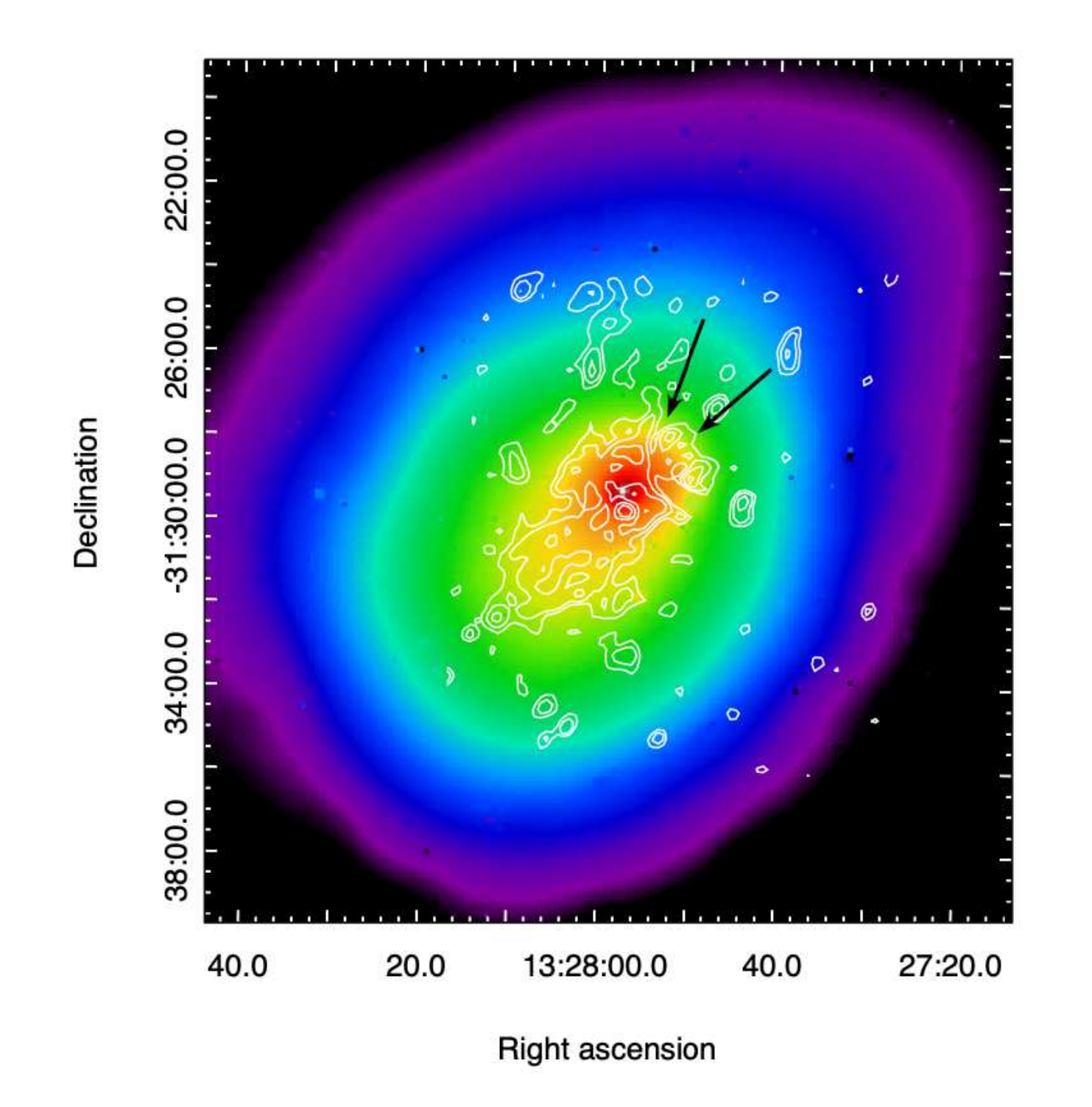}
\vskip 0.2truecm
\includegraphics[scale=0.35]{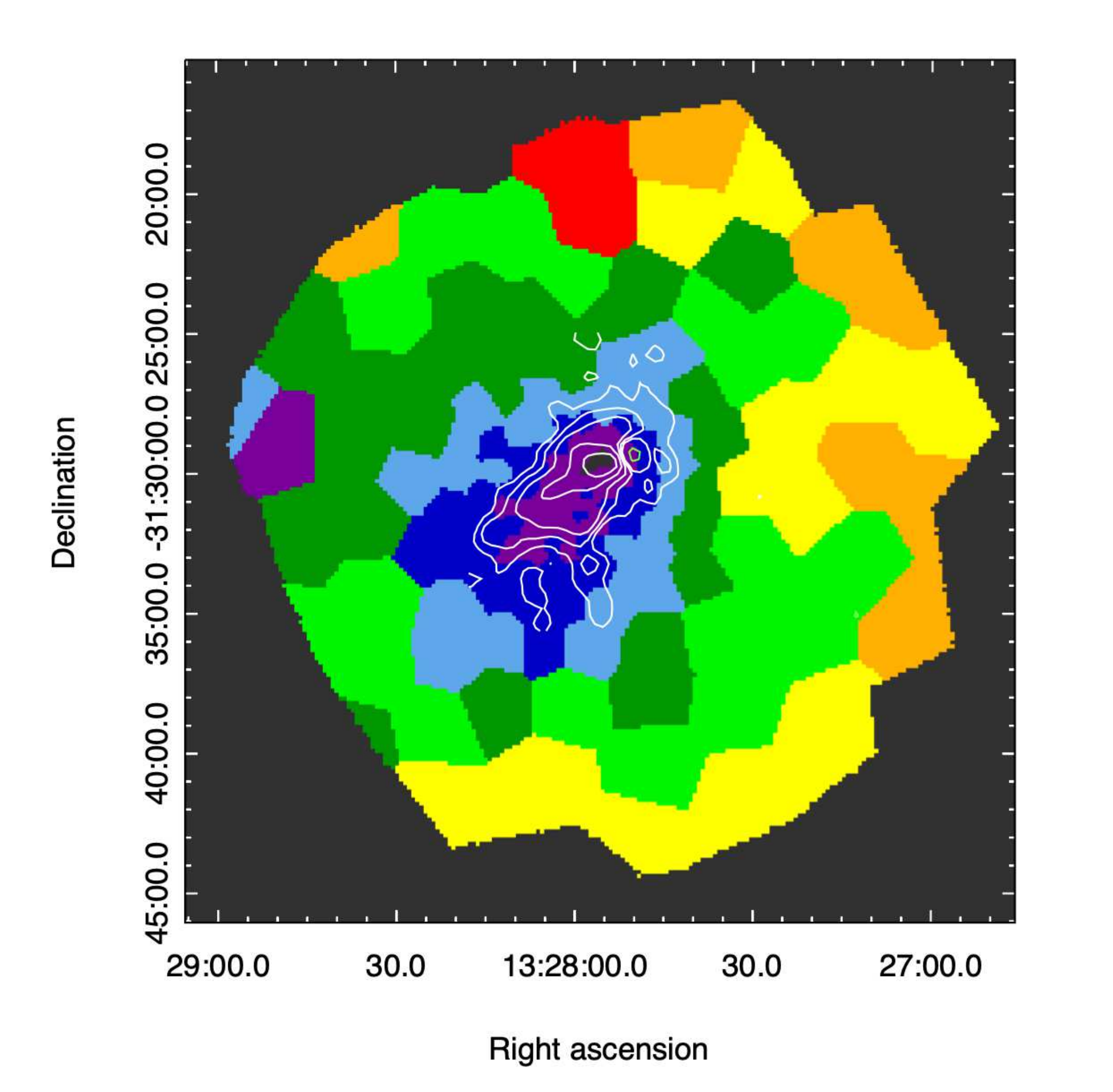}}
\caption{Radio halo in A\,3558. Upper panel:\ ASKAP 887 MHz contours (after subtraction of the
  embedded individual sources; see Sect. 4.7) overlaid on the [0.5,2.5]-keV {\it XMM-Newton}
  image \citep{Rossetti07}. 
  The resolution is
  $25.1^{\prime\prime}\times20.9^{\prime\prime}$, in p.a. $131.5^{\circ}$, and the rms is
  $\sim$35\,$\mu$Jy~beam$^{-1}$. Contour levels start at
  $\pm$0.15 mJy~beam$^{-1}$ and are spaced by $\sqrt2$.  The location of the cold
  front is indicated by the black arrows. 
  In the lower panel, we show MeerKAT 1.283 GHz contours after subtraction of the point sources
  and convolution with a restoring beam of $40.9^{\prime\prime}\times40.4^{\prime\prime}$,
  in p.a. $144.6^{\circ}$, overlaid on the pseudo-entropy (expressed in arbitrary
  units) obtained from the {\it XMM-Newton} observations \citep{Rossetti07}.
  Purple and blue show low pseudo-entropy regions.
  The rms is $\sim$35\,$\mu$Jy~beam$^{-1}$. Contour levels start at
  $\pm$0.125\,mJy~beam$^{-1}$ and are spaced by $\sqrt2$.}
  \label{fig:fig13}
\end{figure}

Diffuse emission was detected  with both ASKAP and MeerKAT. The ASKAP contours of the diffuse emission are shown in the upper panel of Fig.  \ref{fig:fig13} overlaid on the [0.5,2.5-keV] emission imaged by {\it XMM-Newton}. The lower panel shows the MeerKAT contours overlaid on the {\it XMM-Newton} pseudo-entropy map (see Sect. 5.1.3).
The source subtraction in the ASKAP image was carried out using the task SAD in AIPS, down to 0.3 mJy, and the residuals were convolved with a beam of $25.1^{\prime\prime}\times20.9^{\prime\prime}$. This subtraction provided cleaner results in this region than the compact source subtraction methods used earlier. Some residuals are present beyond the bulk of the diffuse emission, however, it is difficult to say whether they are the result of incomplete source subtraction or an indication that the halo could be more extended.
The MeerKAT emission was imaged after subtraction of the individual radio sources in the $u-v$ plane, and the residuals were convolved to a resolution of $40.9^{\prime\prime}\times40.4^{\prime\prime}$. Our 306 MHz and 608 MHz GMRT observations, although they are pointed on the cluster centre, are not sensitive enough to detect this very faint source.

The extent of the diffuse emission is consistent at 887 MHz and at 1.28 GHz, 
that is, $\sim 400\times200$ kpc, and its major axis points towards the small group
SC\,1327--312. The overall shape and size of this structure are consistent
with the brightest region of X--ray emission as detected by {\it XMM-Newton}
\citep{Rossetti07} -- as   discussed in Sect. 5.

We measure a flux density S$_{887~MHz}$=30$\pm3$ mJy and S$_{1283~MHz}$=13$\pm1$ mJy, which lead to a very steep spectrum, $\alpha_{887~MHz}^{1283~MHz}$=\,--2.3$\pm0.4$.
The surface brightness at 1.283 GHz is very low, that is, $\sim$0.1\,$\mu$Jy/arcesc$^2$.
The corresponding radio powers are P$_{\rm 887~MHz}$=1.58$\times10^{23}$\,W~Hz$^{-1}$ and P$_{\rm 1283~MHz}$=6.85$\times10^{22}$\,W~Hz$^{-1}$.

%
%

\section{Discussion}\label{sec:disc}

The ongoing merging and accretion processes in the A\,3558 complex
are supported by observational evidence over the full spectrum, from the
radio to the X--ray, however, their respective details are difficult to disentangle.

Thanks to the unprecedented quality of the radio images presented here, seamlessly covering the frequency range from 0.23 to 1.65 GHz, with the  MeerKAT images reaching $\mu$Jy\,beam$^{-1}$ sensitivity at GHz frequencies over a range of angular scales (from few arcsec to $\sim 10^{\prime}$), our new findings are adding important pieces of information to the details
of accretion and merging in the supercluster core. 

\subsection{Insights from the comparison of the non--thermal and thermal properties}\label{sec:xmmradio}

On the basis of a deep analysis based on {\it XMM-Newton} and {\it Chandra} data, 
\cite{Rossetti07} classified A\,3558 as an intermediate case between cool-core and non cool-core cluster \citep[consistent with the later classification of][]{Hudson10} and argued that most likely the relaxation process of the cluster has been perturbed either by the passage of the small group SC\,1327-312, or by an off-axis merger with a more massive cluster, whose debris are A\,3562 and the two SC groups in the region between A\,3558 and A\,3562.
\cite{Finoguenov04} explained the thermal properties of A\,3562 as due to sloshing induced by the passage of the group SC1329--313 north of A\,3562 itself, coming from east and directed westwards. This is supported by the presence of the ultra-steep spectrum radio halo in A\,3562, whose origin was explained as the result of particle re-acceleration induced by turbulence injected in the ICM by the passage of SC\,1329--313 north of the core of A\,3562 and by the oscillation of the cluster core (sloshing) in the north-south direction (G05).
In the following, we discuss the origin of the various features presented in this paper, going from east (A\,3562) to west (A\,3558), informed by the comparison of X--ray and radio emission as seen in Figs. \ref{fig:fig5}, \ref{fig:fig13}, and \ref{fig:fig14}.

\subsubsection{Region between A~3562 and SC~1329--313}

Figure  \ref{fig:fig5} shows a tight correspondence between the radio and X-ray emission in the whole region going from the centre of A\,3562 and SC\,1329--313.
Beyond the new detection of the arc, both with MeerKAT and {\it XMM-Newton}, from the figure it is clear that the end part of the radio arc (in the west direction) and the bridge connecting A\,3562 and J\,1332--3146a are located in the plateau of X--ray emission connecting A\,3562 and  SC\,1329--313.
Finally, the temperature map shown in Fig. \ref{fig:fig14} has an enhancement in the region between A\,3562 and SC\,1329--313.
From Table~\ref{tab:info}, we infer a mass ratio of $\sim 1:9$ between SC\,1329--313 and A\,3562, namely, we are in the minor merger frame.

The remarkable coincidence of the arc with a similar feature in the X-ray emission lends strong support to the hypothesis made in \cite{Finoguenov04} that SC\,1329--313 has come from the east, travelling north of A\,3562 and then is deflected south to reach its current position.
Such a passage may have induced the sloshing at the origin of the radio halo in A\,3562, as proposed in G05, and injected very mild turbulence in the ICM responsible for the extremely low surface brightness emission of the arc and the bridge.

The observed properties of the radio source J\,1332--3146a are consistent with a merger scenario.
On one hand, we see no evidence of a connection between this diffuse source and the compact radio source associated with the bright member of SC\,1329--313, that is, no radio jets are present at the resolution of our images and the spectral index distribution shows a sharp transition between these two features (see right panel of Fig. \ref{fig:fig8}). On the other hand, the electrons responsible for the diffuse emission may have been deposited in the ICM in a previous cycle of radio activity of the same galaxy and re-accelerated by the turbulence injected in this region by the flyby of SC\,1329--313.  
The steep spectrum of this source and the distribution of the MeerKAT in-band spectral index are consistent with a re-acceleration origin.
We note that the fan-like morphology of the radio halo in A\,3562 and of J\,1332--3146a, opening up towards the north, seems to suggest that the source of the perturbation comes from the north.
\\
Alternatively, the location of J1332--3146a just at the edge of the X-ray emission of J1329--313 (see Fig. \ref{fig:fig5}) opens the possibility of a relic-type source.
The polarisation information, which is not available at present, would be highly valuable with regard to completing the picture of the possible origin of this source.

\medskip
We estimated the equipartition magnetic field in the arc and bridge between A\,3562 and SC\,1329--313, assuming a cylindrical geometry for the region of emission and we obtained the following result: H$_{\rm eq}\sim 0.78$\,$\mu$G. This value is very close to the assumption made in G05 to estimate the strength of the fluid turbulence injected in the ICM in the merger scenario considered here.
As postulated in G05, our results clearly show that indeed the western extension of the radio halo in A\,3562 and J\,1332--3146a are the peaks of a much more extended emission,  which are finally detectable with the current generation of radio interferometers.
\\
Since this emission has thus far only been detected at 1.283 GHz, its spectrum is not yet known. Under the assumption of a spectral break between 0.6 and 1.4 GHz, the corresponding
radiative age of the relativistic electrons responsible for the radio emission ranges from $\sim$132 Myr to 86 Myr (accounting for the losses of the electrons due to the scattering with the cosmic microwave background field H$_{\rm CMB}$). Such values are considerably lower than the estimated age of the interaction between A\,3562 and SC\,1329--313 \citep[$\sim$ 1\,Gyr, see ][]{Finoguenov04}, implying that turbulence has to be active now.

\subsubsection{Region between the two groups SC~1329--313 and SC~1327--312}

The X--ray properties of these two groups were studied with Beppo--SAX \citep{Bardelli02}. No shocks were detected in either of them, which led the
authors to suggest a late merger for the whole structure under study here.

\begin{figure*}[h!]
\centering
    {\includegraphics[scale=0.5]{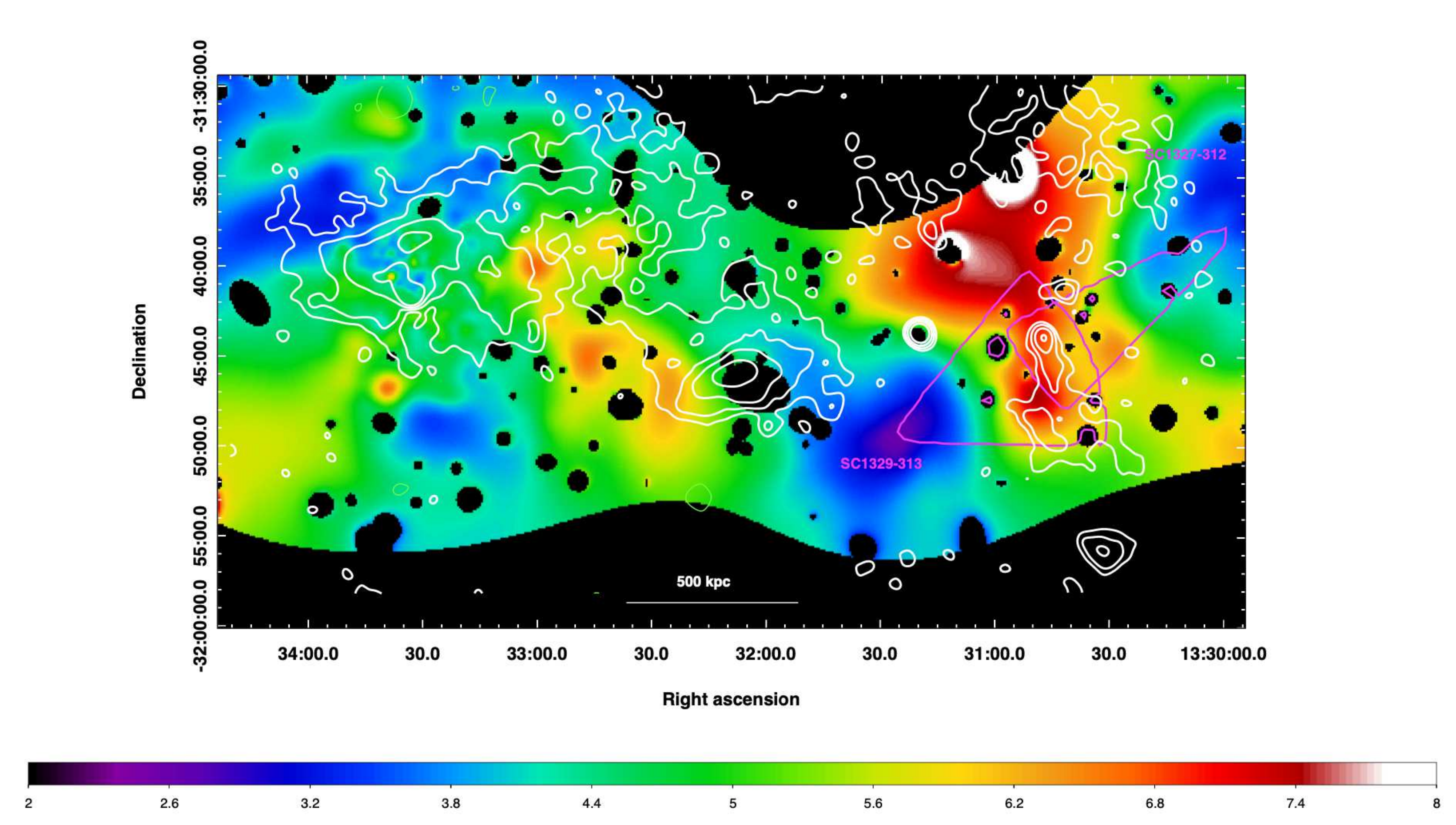}}
    \caption{[0.3,12]-keV {\it XMM-Newton} temperature map (in colour) computed from wavelet
      spectral-imaging
      (see Sect. 3.4). Contours of the  MeerKAT 1.28 GHz image at the resolution of
      $40^{\prime\prime}\times40^{\prime\prime}$ are
      overlaid in white. The Contours are drawn at $\pm$0.1, 0.2, 0.4, 0.8
  mJy\,beam$^{-1}$. The average rms in the image is $\sim 30\mu$Jy\,beam$^{-1}$.
  The black areas
  are the subtracted individual sources. The magenta sectors show the regions of extraction
  of the temperature profiles in SC\,1327--312 and SC\,1329--313 (Sect. 5.2.1 and
  Fig. \ref{fig:fig15}).}
  \label{fig:fig14}
\end{figure*}

\begin{figure}[h!]
{\includegraphics[scale=0.4]{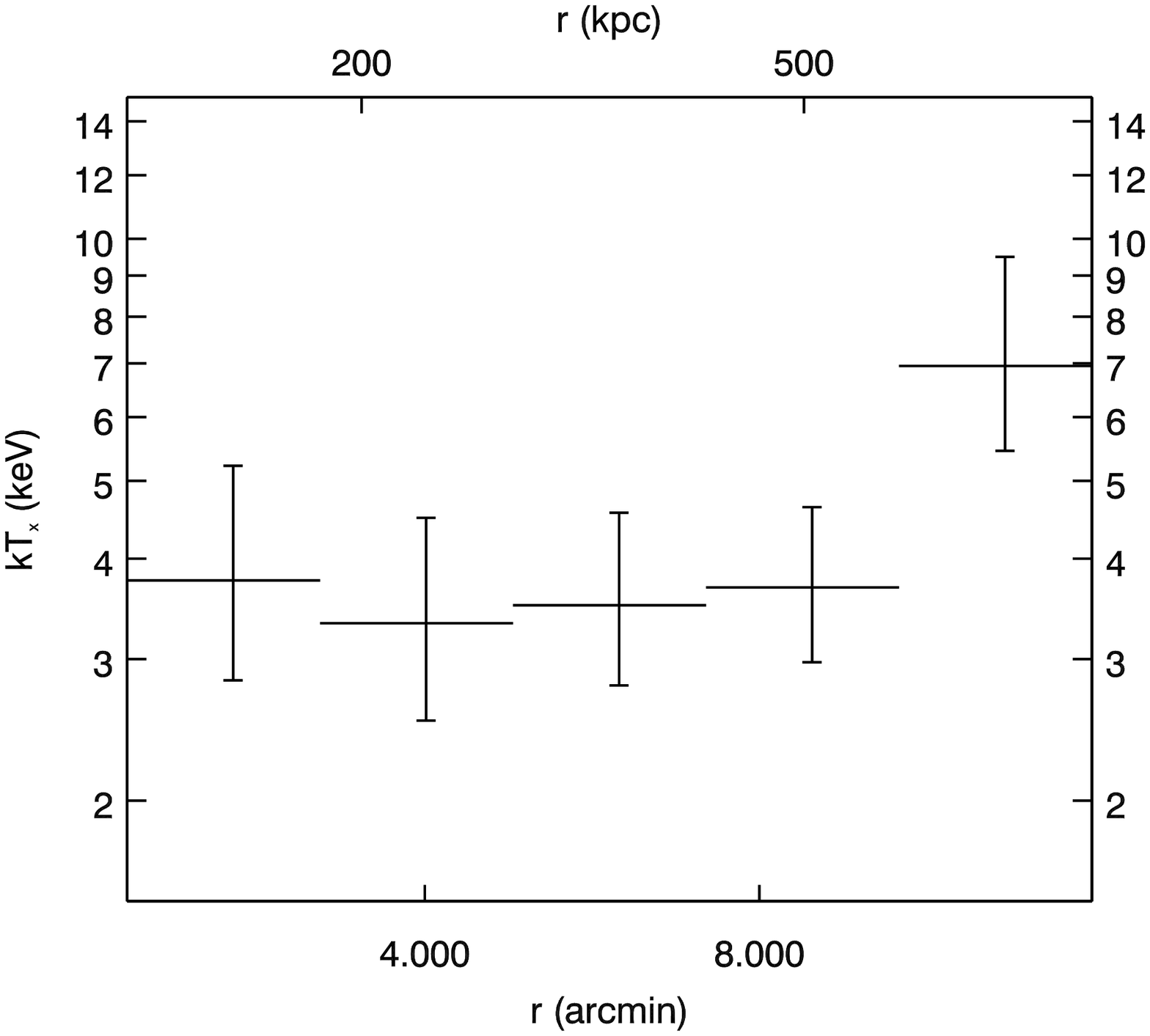}
\includegraphics[scale=0.4]{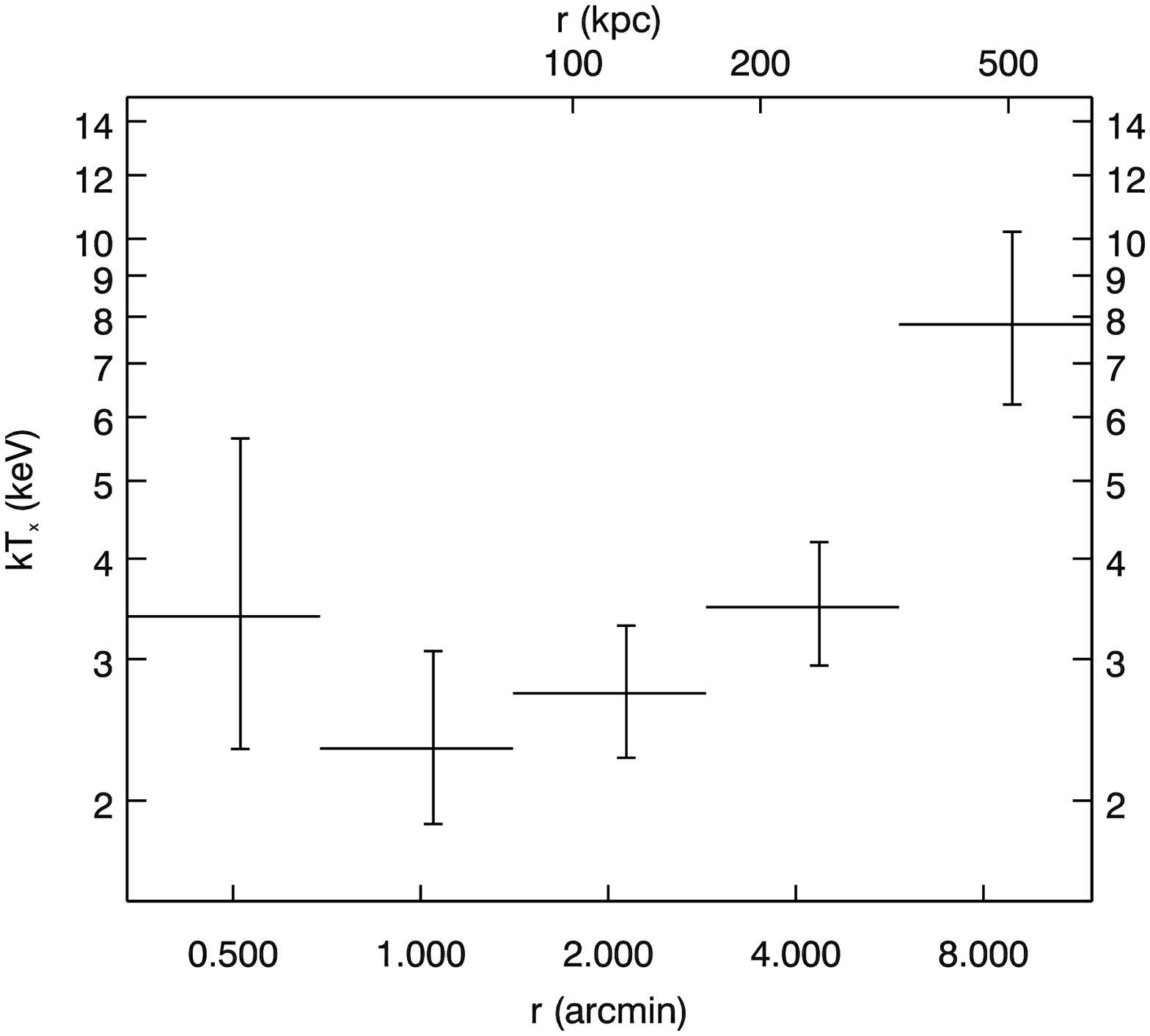}}
    \caption{Temperature profiles derived from the sectors shown in Fig. \ref{fig:fig14}
      for SC\,1327--312 (upper panel) and SC\,1329--313 (lower panel).}
  \label{fig:fig15}
\end{figure}

Figures \ref{fig:fig5} and \ref{fig:fig14} clearly show that the head-tail is located exactly in the bridge of emission connecting the two SC groups, where a temperature enhancement is detected, reaching values of the order of 7-7.5 keV, the highest in the region under investigation. We estimated the significance of this enhancement compared to SC1327--312 and SC\,1329--313 by extracting the temperature profile along the sectors shown in magenta in Fig.  \ref{fig:fig14} over five bins in each sector.
The result is shown in Fig.~\ref{fig:fig15}, with the errorbars corresponding to 1$\sigma$. The significance of the enhancement in the region where the head-tail is located is above 2$\sigma$. Unfortunately, the exposure of the X-ray observations is not deep enough to search for discontinuities in this low surface brightness region.

The combination of this high temperature and the high fractional polarisation in the eastern part of the tail, particularly in the bar (right panel in Fig. \ref{fig:fig12}) are, again, suggestive of interaction between the
radio plasma and some discontinuity.
\\
Tails with bars like the one detected here, are being detected in other clusters, such as Ophiuchus (Giacintucci et al. in prep.) and A\,2443 \citep{Cohen11}.
The morphology and spectral trend (the spectrum steepening towards the bar, which is the steepest feature) for our source bear close similarities to simulations of interactions between radio galaxies and cluster shocks propagating perpendicular to the radio jet axis 
\citep{Nolting19}.
The polarisation properties and the ordered magnetic field which we find would be consistent with this scenario.
The shocks \cite{Nolting19} used in their simulations have Mach numbers in the range between 2-4.  Such strong shocks have not generally been expected in the peripheral regions of a supercluster, but they might be an example of the `runaway' shocks proposed by \cite{Zhang19}, which maintain their strengths in regions with steep density gradients.

The southern emission (Sect. 4.6) lies outside the region covered by the {\it XMM-Newton} observations, as is clear also from Figs. \ref{fig:fig5} and \ref{fig:fig14}. This does not allow any further classification or speculation on its origin.

\subsubsection{Diffuse emission at the centre of A~3558}

The upper panel of Fig. \ref{fig:fig13} clearly shows that the diffuse source at the centre of A\,3558 is perfectly coincident with the brightest part of the  X--ray emission as imaged by
{\it XMM-Newton}.

A\,3558 is an intriguing cluster.
Based on the analysis of {\it XMM-Newton} and {\it Chandra} observations, \cite{Rossetti07}  concluded that the cluster has properties both of cool-core and merging clusters, with a cold front in the north-western edge (whose position is shown in the upper panel of Fig. \ref{fig:fig13}) and a low entropy tail (shown in colour in the lower panel of Fig. \ref{fig:fig13}), but with an overall lack of central symmetry in its thermal properties.
A multi-wavelength study of the brightest cluster galaxies (BCG) in the core of the Shapley Supercluster \citep{diGennaro18} reports that the BCG in A\,3558 has radio properties that are similar to those in merging clusters.
The diffuse radio emission detected at its centre (Sect. 4.7) reflects this anomalous situation.
The extent of the radio emission is smaller than typical radio halos ($\sim 400\times200$ kpc), and its power, P$_{\rm 1.283~GHz}=6.85\times10^{22}$ W\,Hz$^{-1}$, is very low.
If we consider the value of M$_{\rm 500}$ given in Table \ref{tab:info} and scale
the radio power to 1.4 GHz using the spectral index obtained here ($\alpha$=--2.3), we see that this
this emission is extremely underluminous compared to what is expected from
the M$_{\rm 500}$-P$_{\rm 1.4~GHz}$ correlation for radio halos \citep{Cuciti21,Duchesne21b}.
On the other hand, the emission is in reasonable agreement with the L$_{\rm X}$-P$_{\rm 1.4~GHz}$ correlation for mini-halos \citep{Kale15}.
We note that its north-western boundary is coincident with the cold front detected in \cite{Rossetti07} and highlighted in the upper panel of Fig. \ref{fig:fig13} and the overall emission is aligned along the bright ridge of X-ray emission.
This region corresponds to the low-entropy tail (see lower panel of Fig. \ref{fig:fig13}).

\cite{Rossetti07} explained the thermal properties of A\,3558 as due to sloshing induced by a perturber, most likely SC1327--312. The underluminous very steep spectrum of the diffuse emission in A\,3558 is consistent with the possibility that the origin of this source is sloshing induced by a minor merger, the mass ratio between SC1327--312 and A\,3558 being $\sim 1:5$.
The case of A\,3558 is reminiscent of  A\,2142, another cluster with intermediate properties between cool-core and non-cool-core clusters \citep{Rossetti13} where large-scale sloshing was invoked to account for the origin of the two-component Mpc-scale radio halo \citep{Venturi17b} Whether this is a radio halo or a mini-halo remains unclear, and this sharp classification is probably inadequate to reflect the complexity of the Shapley Supercluster core and of the relation between radio halos and cluster mergers at a more general level. This is also shown in \cite{Savini19}, where large-scale radio emission in the form of radio halos in non-merging clusters has been detected at low frequencies.

%
%

%
%

\subsection{Origin of the radio tail of SOS~61086}\label{sec:radioopt}

\begin{figure}[h!]
\centering
\includegraphics[scale=0.4]{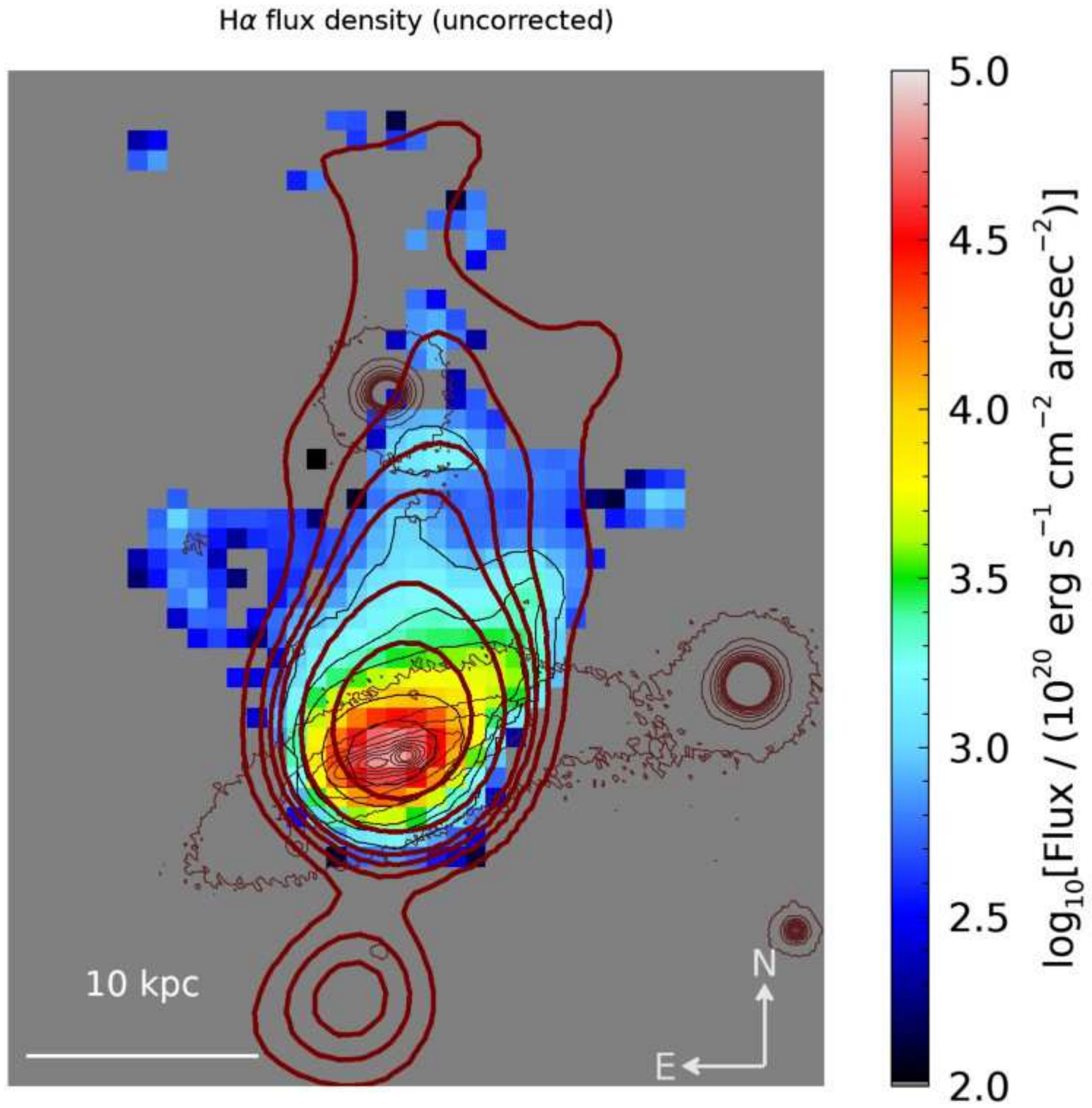}
\caption{H$\alpha$ flux for SOS\,61086 derived from IFS data. The $r$-band contours and the
  MeerKAT radio contours (same as Fig. \ref{fig:fig9}, left panel) are shown
  in thin and thick red lines, respectively.}
  \label{fig:fig16}
\end{figure}

A fundamental piece of evidence of the ongoing RPS affecting the galaxy SOS\,61086 has been provided by the distribution of the H$\alpha$ emission derived from IFS and shown in Fig.~\ref{fig:fig16}. The IFS data indicate that ionised gas spreads out from the disk of the galaxy in a sort of triangular region with a vertex in the central disk and one side at $\sim$16\,kpc north and directed approximately E-W with clumps of gas extending further in the north, reaching $\sim$30\,kpc in projection. The gas disk appears clearly truncated beyond $\sim$6\,kpc from the centre along the major axis. A possible interaction with the other member of SC\,1329-313 located $\sim 17$\,kpc north of SOS\,61086 (see left panel of Fig.~\ref{fig:fig9}) has been excluded considering the lack of any sign of perturbation in the stellar disk of this `red and dead' galaxy, as well as the shape and extent of the ionised gas tail going beyond the northern galaxy in projection.


In SOS\,61086, the radio-emitting plasma and the ionised gas show equal
extent in projection from the galaxy disk (see
Fig.~\ref{fig:fig16}), although the ionised gas tail is wider
($\sim 28$\,kpc E-W wide at $\sim 16$\,kpc North of the disk) and it
looks fragmented into clumps in the section farthest from the disk. We
note that the H$\alpha$ flux in Fig.~\ref{fig:fig16} is above 5$\sigma$
level\footnote{ The sensitivity of the IFS observations was $0.5\times 10^{-17}\,\mathrm{erg}\,\mathrm{cm}^{-2}\,\mathrm{s}^{-1}\,\mathrm{arcsec}^{-2}$ at a signal-to-noise ratio SNR=5 for the H$\alpha$ line \citep{Merluzzi16}.} almost everywhere, and the first radio continuum contour in the tail corresponds to $\sim$3.5 times the local noise.
In addition, the two emissions overlap in the galaxy disk, suggesting that the ram pressure causing the truncation of the gas disk affects the radio-emitting plasma with similar efficiency. Therefore, we can confirm that, at our sensitivity, the magnetic field and the warm ionised gas traced by the H$\alpha$ emission are closely linked, in agreement with what is observed in the Virgo galaxy
NGC\,4438 \citep{Vollmer09}.

Ram-pressure stripped tails detected in radio continuum have been
identified and analyzed in the Virgo cluster \citep{Vollmer04,Vollmer09}, the Coma cluster \citep{Crowl05} and  Abell\,1367 \citep[see][and references therein]{Gavazzi95}.
Recently, \cite{Chen20} carried out a 1.4\,GHz continuum and H{\sc i} emission survey in the Coma cluster to investigate the radio properties of RPS galaxies. They detected radio continuum tails in 50\% of the targets demonstrating the widespread presence of relativistic electrons and magnetic fields in the stripped tails. Now, the origin of the radio continuum emission in the tail of
each RPS galaxy is under debate, assumed to be possibly related either to 'in situ' star formation \citep[][and references therein]{Gavazzi95} or to relativistic electrons stripped from the galaxy by ram pressure and possibly re-accelerated \citep[i.e.,][]{Chen20}.
In order to distinguish and quantify these two contributions, multi-band data are mandatory, as shown in the above studies. This approach is also essential since different physical processes may contribute to the emission at each wavelength. The ionised gas as traced by the H$\alpha$ radiation, for instance, can be photoionised in the H{\sc ii} regions, namely, it maps the youngest stellar population, or it may be shock-ionised in the turbulent tail by the ram pressure \citep[i.e.,][]{Merluzzi13}.

\subsection{Multi-band analysis of the galaxy extraplanar emission}

Beside the morphological conformity, for SOS\,61068 the other pieces of evidence supporting a scenario where the radio-emitting plasma and warm gas tails originate from the same mechanism, that is, RPS, is given by the consistency (see Sect. 4.4 and references therein) among the epoch of the onset of stripping ($\sim 250$\,Myr), the age of the youngest stellar population ($< 200$\,Myr), and the age of the radio tail ($\sim 100$\,Myr).

For SOS\,61086, the line ratios observed in the ionised gas indicate that it is mainly photoionised by hot stars, both in the galaxy disk and in the tail, and that shocks are important in determining the excitation only in a NE region of the tail. The integrated H$\alpha$-derived star formation rate (SFR) of SOS\,61086, excluding the contribution of the shock-ionised gas, amounts to $1.8\pm 0.6$\,M$_\odot$yr$^{-1}$, half of which occurs in a central region of 2.5\,kpc radius \citep{Merluzzi16} and with lower star formation in the detached gas.  
\cite{Haines11} derived the relation between the SFR and the 1.4\,GHz flux for the Shapley galaxies. Considering the SFR and the measured 1.4\,GHz flux of SOS\,61086 (1.34\,mJy), the galaxy follows the sequence of the star-forming spiral galaxies and does not show any radio excess (see Fig.~5 of \cite{Haines11}).

All together, this evidence points towards a common cause of the radio and H$\alpha$ tails, but does not enable us to disentangle the origin of the relativistic electrons in the tail. The relative lengths and strength of the radio continuum emission and the H$\alpha$ extraplanar tails may help in this investigation. \cite{Gavazzi95} observed two cases where the radio tail significantly exceeds the H$\alpha$ one in projection. By contrast, the radio continuum tails detected in the RPS galaxies of the Coma cluster by \cite{Chen20} are usually shorter than the H$\alpha$ tails. Of course, these differences in the relative extents of the radio and H$\alpha$ tails can be explained with differences in sensitivities among the studies, but with respect to our data and those of \cite{Chen20}, this seems not to be the case. Both radio observations and H$\alpha$ fluxes reach comparable depths and the projected lengths of the radio and H$\alpha$ tails are similar. 
Considering the age estimated for the relativistic electrons ($\le 100$\,Myr) and the projected length of the tail ($\sim 30$\,kpc) we obtain a velocity of the gas stripped from the galaxy of $\sim 300$\,km\,s$^{-1}$. This value is actually a lower limit because we are using a projected distance and an upper limit for the age of the relativistic electrons and, thus, it is consistent with the wind velocity ($V_{wind}=750$\,km\,s$^{-1}$) inferred from the N-body/hydrodynamical simulations of RPS run for this particular case \citep[see][]{Merluzzi16}. The result of this simple calculation supports the scenario where the radio emission in the tail is fuelled by the relativistic electrons stripped from the galaxy.

The other parameter that informs us about the nature of the radio tail is the spectral index. \cite{Vollmer09} found that the magnetic field and the warm ionised gas traced by H$\alpha$ emission are closely linked in the Virgo galaxy NGC\,4438 and that the spectral index of the extraplanar radio emission does not steepen with increasing distance from the galaxy disk implying an in situ re-acceleration of the relativistic electrons. This is different from
what is observed in NGC\,4522 \citep{Vollmer04} and NGC\,4569 \citep{Chyzy06}, with the relativistic electrons showing a rapid aging with increasing distance. A steepening of the spectral index has been also reported by \cite{Chen20} \citep[see also][]{Muller21}.
The MeerKAT in-band spectral index shows that even the tail SOS\,61086 steepens away from the location of the galaxy, reaching values up to $\alpha\sim$\,--2 at the end of the bright portion of the tail (see right panel of Fig. \ref{fig:fig9} and inset in the right panel of Fig. \ref{fig:fig8}), consistent with synchrotron cooling, as expected for a radio emission due to relativistic electrons stripped from the galaxy.

We conclude that the radio tail revealed by MeerKAT observations in SOS\,61086 is due to the ongoing RPS and is mainly related to relativistic
electrons stripped from the galaxy disk.

SOS\,61086 provides the observational evidence that RPS can also
affect low mass galaxies in moderate- or low-density environment \citep{Marcolini03,Roediger05}.
Moreover, it shows that ram pressure may act quite efficiently even when the galaxy orbit is not radial, with the stripped tail oriented almost tangentially to the cluster SC\,1329--313. This feature can be easily accommodated in the scenario suggested here: the minor merging between SC\,1329--313 and A\,3562 would form the bridge of low surface brightness emission between the two clusters and further perturb the ICM \citep[i.e.][]{Owers12},
triggering RPS in SOS\,61086.

Very recently, \cite{Roberts21} identified 95 star forming galaxies experiencing ongoing RPS in low redshift (z$<$0.05) clusters through 120-168\,MHz radio continuum images from the LOFAR Two-meter Sky Survey. Hopefully, these data will shed light on the possible origin of the radio continuum emission in the tail of RPS galaxies.

\section{Conclusions}\label{sec:conc}

In this paper, we present a radio study of the central region of the Shapley Supercluster, encompassing the two Abell clusters A\,3558 and A\,3562 and the two SC groups between them, SC\,1327--312 and SC\,1329--313. Our observations were carried out with ASKAP, MeerKAT, and GMRT, in a frequency range from 233 MHz to 1.656 GHz, and our analysis was complemented with {\it XMM-Newton} information and ESO-VST optical data.

Beyond the radio halo in A\,3562 and the radio source J\,1332--3146a, already imaged in previous works (V03, G05, V17), we have detected several new features. In particular:

\begin{itemize}
  
\item{} We revealed diffuse emission on the supercluster scale, covering the whole region between A~3562 and SC\,1329--313, corresponding to an extent of $\sim$\,1 Mpc.
The emission has extremely low surface brightness, $\sim$ 0.09\,$\mu$Jy/arcsec$^2$.
It is the first time that diffuse emission between a cluster and a group is detected at GHz frequencies.
It has the shape of a bridge connecting the radio halo in A\,3562 and SC\,1329--313, plus a well-defined arc connecting the cluster and the SC group from the north.
The equipartition magnetic field of this emission is very low,  H$_{\rm eq}\sim$0.78\,$\mu$G, and the estimated age of the radiating electrons range between $\sim$86 Myr to $\sim$132 Myr depending on the assumptions on the spectral break. Such a timescale is much shorter than the scale of the merger, which is estimated to be 1 Gyr;

\item{} The diffuse radio source J1332--3146a shows no hints of connection with the embedded compact radio emission associated with a bright galaxy in SC1329--313. Its spectral properties are consistent with a re-acceleration origin;

\item{} We identified a radio continuum tail induced by ram pressure affecting the galaxy SOS\,61086 in SC\,1329--313. This case shows that ram pressure stripping can involve both warm gas and radio-emitting plasma and highlights the role of cluster-cluster interaction in triggering it;

\item{} A head-tail radio galaxy has been identified, whose tail is broken and culminates in a misaligned bar, as is now being observed in a number of clusters. The broken tail and the bar have a very steep spectrum. A preliminary analysis shows that the bar has very high fractional polarisation, which deserves further investigation;

\item{} We detected diffuse radio emission at the centre of A\,3558 for the first time.
The radio source is small in size compared to typical giant radio halos, at 400$\times$200 kpc, and it has an ultra steep spectrum, $\alpha_{\rm 887 MHz}^{\rm 1283 MHz}=-2.3\pm0.4,$ and a very low radio power, P$_{\rm 1283 MHz}=6.85\times10^{22}$ W\,Hz$^{-1}$.

\end{itemize}

A comparison of our radio images with the X--ray emission and temperature map from {\it XMM-Newton} shows a tight correspondence between the arc and bridge of radio emission and similar features in the X-rays.

The radio arc and the bridge support the scenario of \cite{Finoguenov04}, who suggested an off-axis interaction between A\,3562 and SC\,1329--313, the latter passing north of A\,3562 to reach its current location.
The overall morphology of the radio halo in A\,3562, J\,1332--3146a and the radio tail of SOS\,60861 are consistent with a perturbation coming from the north.
We propose that the arc and the bridge could trace the channel of turbulence injected in the ICM  by passage of SC1329--313. The value we derived for the equipartition magnetic field is consistent with the estimates provided in G05 to support the turbulence needed to produce such emission.

The classification of the diffuse radio emission in A\,3558 is unclear, and this possibly reflects the fact that the cluster shows features of both cool-core and non-cool-core systems. Its overall radio properties, namely, its ultrasteep spectrum, small size, and extremely low radio power, are consistent with re-acceleration induced by sloshing. In addition, these details would  lend support to the interpretation of the X--ray properties of the cluster as being due to sloshing induced by a perturber -- most likely SC\,1327--312 \cite{Rossetti07}.

The location of the head-tail and bar in the ridge of X-ray emission just between SC\,1327--312 and SC\,1329--313, a region of enhanced temperature, coupled with the steep spectrum and high fractional polarisation in the bar, suggest interaction between these two low-mass SC groups.
Unfortunately, the available X-ray observations are not deep enough to detect discontinuities in the ICM.
 
Our study shows that even minor mergers may leave detectable non-thermal signatures, such as Mpc-scale radio emission in the region between clusters and groups, as well as other extended features associated with early- and late-type cluster galaxies, which bear invaluable information on the formation of large-scale structures. Deeper X-ray and polarisation observations have been planned to further advance our understanding of the merger and accretion processes in the core of the Shapley Supercluster.

\medskip\noindent
\begin{acknowledgements}
We thank the anonymous referee for helping the clarity of the paper. The Australian SKA Pathfinder is part of the Australia Telescope National Facility which is managed by CSIRO. Operation of ASKAP is funded by the Australian Government with support from the National Collaborative Research Infrastructure Strategy. ASKAP uses the resources of the Pawsey Supercomputing Centre. Establishment of ASKAP, the Murchison Radio-astronomy Observatory and the Pawsey Supercomputing Centre are initiatives of the Australian Government, with support from the Government of Western Australia and the Science and Industry Endowment Fund. We acknowledge the Wajarri Yamatji people as the traditional owners of the Observatory site.
The MeerKAT telescope is operated by the South African Radio Astronomy Observatory (SARAO), which is a facility of the National Research Foundation, an agency of the Department of Science and Innovation. The National Radio Astronomy Observatory is a facility of the National Science Foundation operated by Associated Universities, Inc.
The optical imaging is collected at the VLT Survey Telescope using the Italian INAF Guaranteed Time Observations.
We thank the staff of the GMRT who made these observations possible. The GMRT is run by the Nationale Centre for Radio Astrophysics of the Tata Institute of Fundamental Research.
T. Venturi and G. Bernardi acknowledge the support from the Ministero degli Affari Esteri e della Cooperazione Internazionale, Direzione Generale per la Promozione del Sistema Paese, Progetto di Grande Rilevanza ZA18GR02. R. Kale acknowledges the support of the Department of Atomic Energy, Government of India under project no. 12-R\&D-TFR-5.02-0700. S.\,P. Sikhosana acknowledges funding support from the South African Radio Astronomy Observatory (SARAO) and the National Research Foundation (NRF).
O. Smirnov's research is supported by the South African Research Chairs Initiative of the Department of Science and Technology and National Research Foundation.
\end{acknowledgements}
\bibliographystyle{aa}
\bibliography{venturi.bib}
\begin{appendix}
\section{GMRT Images}     

In this appendix, we show the primary beam corrected images of the GMRT observations presented
in Table~\ref{tab:logs}. The location of the various features studied in this paper are
highlighted in each image. Since we did not use the observations at 608 MHz of project
22\_039 in this paper, we do not show those images.

%
\begin{figure*}[h!]
  \vskip 0.2truecm
\includegraphics[scale=0.77]{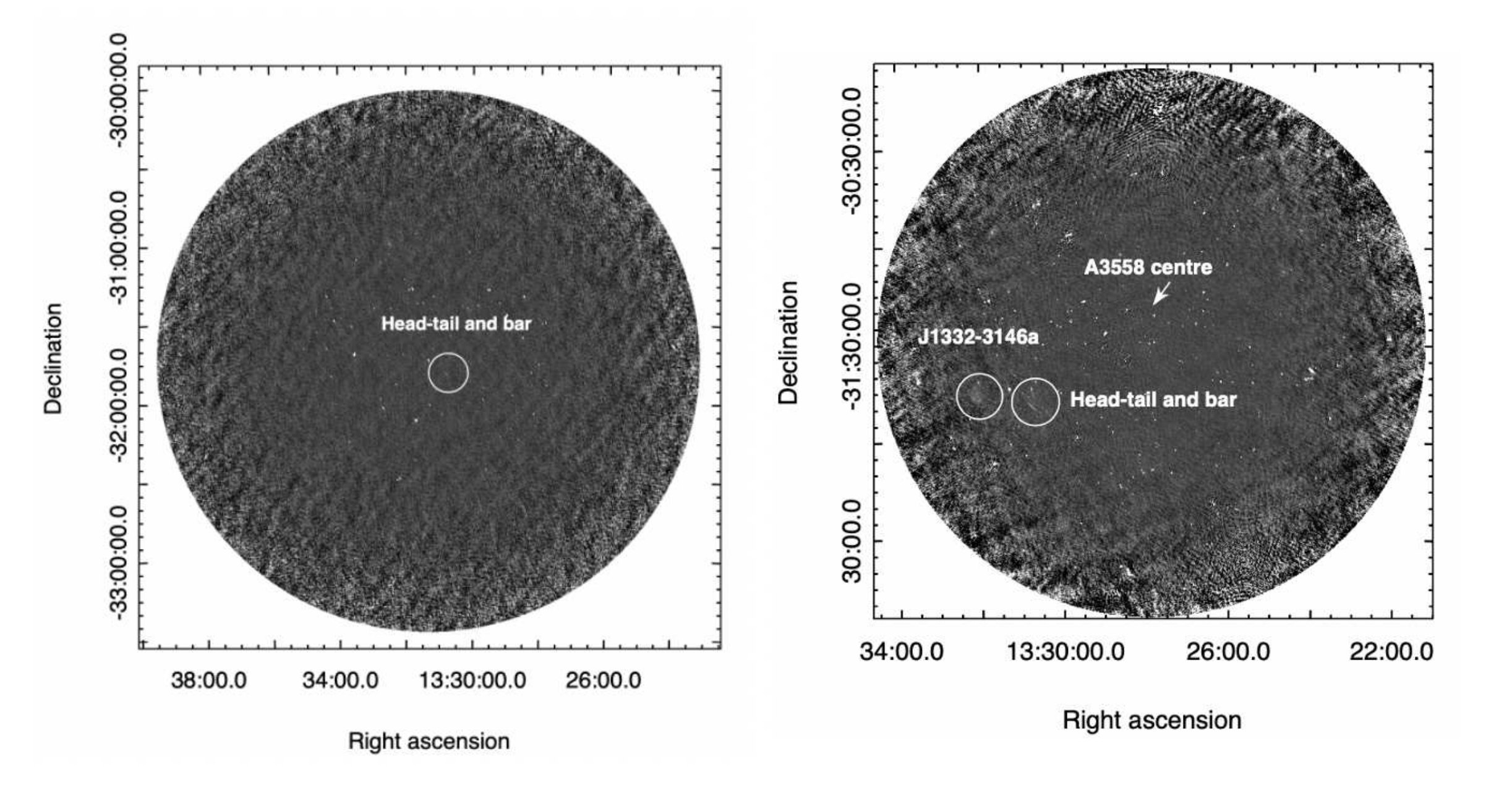}
  \caption{Primary beam corrected 233 MHz GMRT image of the project 30\_024 (see
  Table~\ref{tab:logs} and Fig.\ref{fig:fig2}), shown in the left panel. The angular resolution is
  $24.4^{\prime\prime} \times 10.7^{\prime\prime}$, p.a. $26^\circ$. The location of the 
  head-tail and the bar has been highlighted. The right panel shows the primary beam corrected 306 MHz GMRT
  image of the project 22\_039 (see Table~\ref{tab:logs} and Fig.\ref{fig:fig2}).
  The angular resolution is
  $14.0^{\prime\prime} \times 9.5^{\prime\prime}$, p.a. $18.4^\circ$.  The location of
  J\,1332--3146a, head-tail, bar, and A\,3558 centre have been highlighted.}
  \label{app-fig1-2}
\end{figure*}
%
\begin{figure*}[h!]
\vskip 0.2truecm
\includegraphics[scale=0.77]{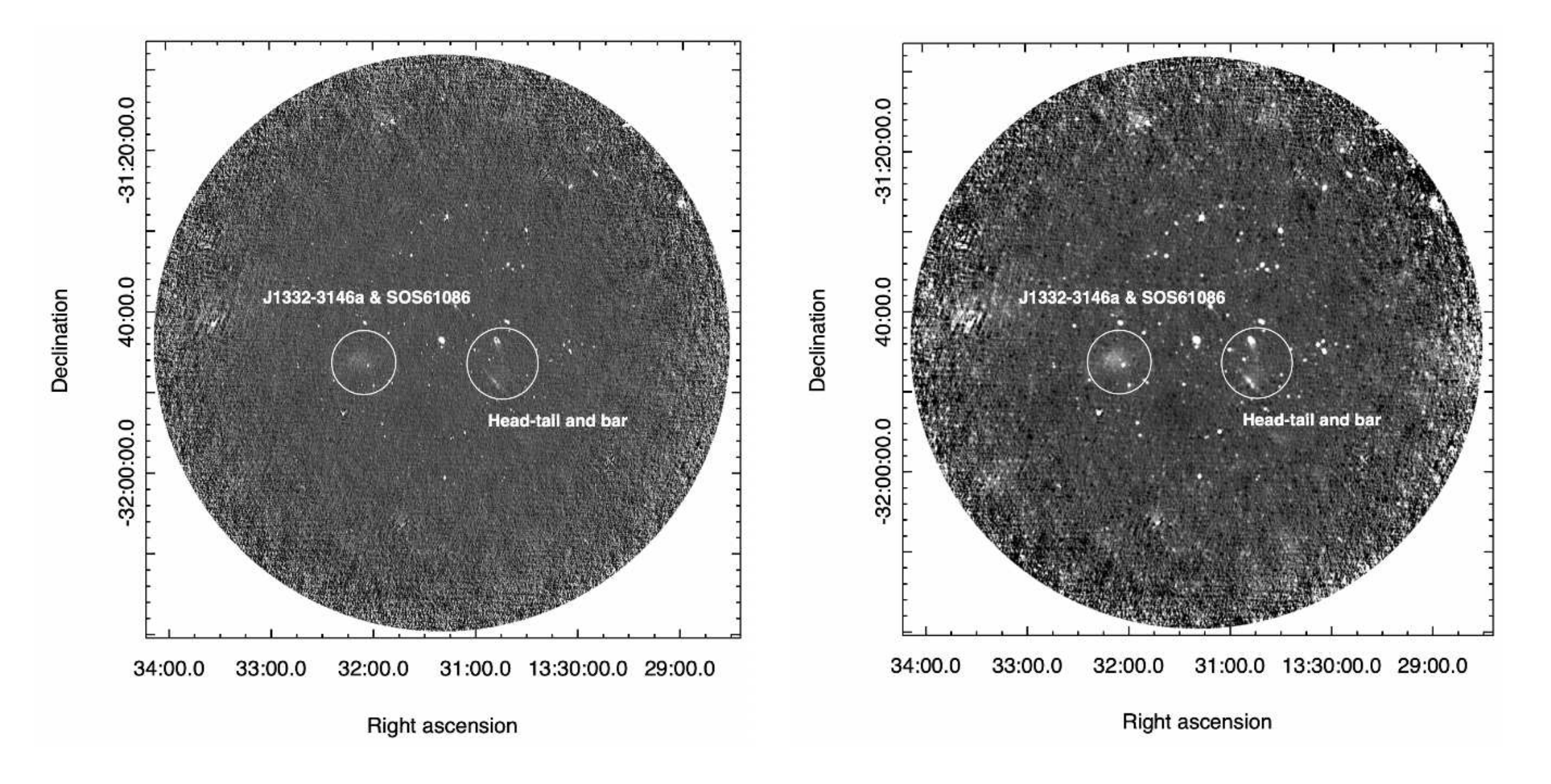}
\caption{Primary beam corrected 607 MHz GMRT images of the project 30\_024 (see
    Table~\ref{tab:logs}, Fig. \ref{fig:fig2}, and Sect. 4.5).
    Location of
    J\,1332--3146a, SOS\,61086, head-tail, and bar have been highlighted.
    Left panel:\ Angular resolution is $9.7^{\prime\prime} \times 5.5^{\prime\prime}$,
    p.a. $9.6^\circ$. Right panel: Angular resolution:
  $15^{\prime\prime} \times 15^{\prime\prime}$.}
  \label{app-fig3-4}
\end{figure*}
\end{appendix}
\end{document}